\documentclass[11pt]{article}

\usepackage{amsthm}
\usepackage{amsmath}
\usepackage{amssymb}
\usepackage{amsfonts}
\usepackage{epsfig}
\usepackage{graphicx}
\usepackage{color}
\usepackage{float}
\usepackage[caption = false]{subfig}
\usepackage{algorithm,algorithmic}
\usepackage{multirow}

\newtheorem{theorem}{Theorem}

\newtheorem{lemma}[theorem]{Lemma}

\allowdisplaybreaks

\DeclareMathOperator*{\argmax}{arg\,max}
\begin{document}
\title{\bf Robust and probabilistic optimization of dose schedules in radiotherapy}
\author{Hamidreza Badri${}^{1}$, Yoichi Watanabe ${}^{2}$, Kevin Leder${}^{1}$\\
{\small 1. Department of Industrial and Systems Engineering}\\{\small 2. Department of Radiation Oncology}\\
{\small University of Minnesota, Minneapolis, MN} \\
\date{\today} }

\normalsize{}
\maketitle

\begin{abstract}
We consider the effects of parameter uncertainty on the optimal radiation schedule in the context of the linear-quadratic model.
Our interest arises from the observation that if inter-patient variations in normal tissues and tumor sensitivities to radiation or sparing factor of the organ at risk (OAR) are not accounted for during radiation scheduling, the performance of the therapy may be strongly degraded or the OAR may receive a substantially larger dose than the maximum threshold. This paper proposes two radiation scheduling concepts to incorporate inter-patient variability into the scheduling optimization problem. The first approach is a robust formulation that formulates the
problem as a conservative model that optimizes the worst case dose scheduling that may occur, assuming that the
parameters vary within given intervals. The second method is a probabilistic approach, where the model parameters are given by a set of random variables. Our probabilistic formulation insures that our constraints are satisfied with a given probability, and that our objective function achieves a desired level with a stated probability. We used the same transformation as \cite{SaGhKi15} to reduce the resulting optimization problem to two dimensions. We showed that the optimal solution in the absence of uncertainty in the tumor radio-sensitivity parameters ($\alpha$ and $\beta$) occurs at one of the corners of the feasible region. However if we incorporate uncertainty in $\alpha$ and $\beta$ into the optimization problem, this result does not hold anymore. In this case, we showed that the optimal solution lies on the boundary of the feasible region and we implemented a branch and bound algorithm to find the global optimal solution. We demonstrated how the configuration of optimal schedules in the presence of uncertainty compares to optimal schedules in the absence of uncertainty (conventional schedule).  We observed that if the number of fractions in the optimal conventional schedule is the same as the robust and stochastic solutions, it is preferable to administer equal or smaller total dose. In addition if there exist more (fewer) treatment sessions in the probabilistic or robust solution compared to the conventional schedule, a reduction in total dose squared (total dose) will be expected.  Finally, we performed numerical experiments in the setting of head-and-neck tumors including several normal tissues to reveal the effect of parameter uncertainty on optimal schedules and to evaluate the sensitivity of the model to the choice of key model parameters.

 
\end{abstract}
 
 \section{Introduction}
 
The building block for virtually all mathematical models of radiation response is the linear-quadratic model (LQ), which matches well with experimental data across a wide range of clinically relevant radiation doses and fractionation schemes (\cite{fowler1989linear} and \cite{brenner2008linear}). The basic model states that if a collection of cells is exposed to $N$ fractions of radiation, $d_i$ Gy (SI derived unit of ionizing radiation) in $i^{th}$ fraction, the reproductively viable fraction of cells after the exposure is given by $e^{-\sum_{i=1}^{N}\alpha d_i+\beta d_i^2}$. The two parameters $\alpha$ and $\beta$ depend on the specific tissue that is being irradiated. The parameter $\alpha$ represents killing of cells from a single track of radiation, and $\beta$ represents the killing of a cell via two independent tracks of radiation \cite{hall2006radiobiology}. There are several mathematical extensions to the LQ framework to incorporate additional biological phenomena such as repopulation of the tumor population between fractions, re-oxygenation of the tumor (this is required for some radiation therapy to be effective), the effectiveness of DNA repair mechanisms between fractions, and the redistribution of tumor cells within the cell cycle. Taken together these four extensions are often referred to as the `4Rs' and there have been several works based on these extensions \cite{withers1975four}. 

When radiotherapy is used in the clinical setting, it is necessary to ensure that the treatment avoids excessive toxicity in normal tissues in the vicinity of the tumor. Therefore it is necessary to ensure that the radiation absorbed by the surrounding normal tissue falls within desired constraints.
Hence the ultimate goal in radiotherapy is maximizing tumor damage while ensuring that the level of normal tissue toxicity does not exceed a given threshold. The standard approach for measuring tumor damage and tissue toxicity is via the linear quadratic model and the biologically equivalent dose (BED), respectively (\cite{fowler1989linear} and \cite{fowler201421}).

Most radiation treatments are currently administered in equal fractions five days a week, for 6 weeks total. Over the past few decades there have been several mathematical works that have studied the survival benefit of various fractionation schedules for a wide range of cancers.
In \cite{brenner1998linear}, that most other radiobiological make similar time-dose predictions as the LQ formalism. In that work they used the LQ model in combination with Lea-Catcheside time factor, which takes into account dose protraction or fractionation and DNA repair between fractions. Yang and Xing (\cite{yang2005optimization}) explored the influence of the `4Rs' of radiobiology on external beam radiotherapy for fast and slowly proliferating tumors and conclude that including repair effects in the BED model may give rise to optimal non-uniform fractionation schedules. Mizuta et al. (\cite{mizuta2012mathematical}) presented a mathematical model that minimizes the radiation effect on the late responding normal tissues while keeping the effect of radiation on the tumor constant.  They showed that the multi-fractionated irradiation with a constant dose is better if the ratio of $\left( \frac{\alpha}{\beta}\right)_{\text{Normal Tissue}}/\left( \frac{\alpha}{\beta}\right)_{\text{Tumor}}$ is less than the ratio of the dose received by the normal tissue, while Hypo-Fractionated irradiation is better otherwise. Unkelbach et al. (\cite{unkelbach2013dependence}) studied the interdependence of the optimal fractionation scheme and the spatial dose distribution in the normal tissues. In particular, they derived a criterion under which a Hypo-Fractionated regimen is indicated for both parallel and serial OARs. In a very recent work \cite{saberian2014optimal}, a formulation of the optimal fractionation problem that includes multiple normal tissues has been considered. They established sufficient conditions under which equal-dosage or single-dosage fractionation is optimal. In recent work (\cite{leder2014mathematical}) the authors investigated optimal fractionation for a mouse model of glioblastoma, in this work they found that non-standard fractionation schedules lead to improved survival times; a finding that was verified in experimental studies. In \cite{badri} this work was extended to include a richer set of toxicity constraints.

Until very recently, there was no work that precisely described the optimal fractionation sizes in the presence of multiple normal tissues. In particular, most works considered the optimal schedule with respect to a single normal tissue. However in practice, there are usually at least two healthy structures in the vicinity of the tumor. Saberian et al. considered several normal tissues in their study, however they were unable to find the closed form solution to the problem for all possible cases and they only discussed the sufficient conditions under which equal-dosage fractionation is optimal  \cite{saberian2014optimal}. In \cite{badri} two simultaneous normal tissue toxicity constraints were implemented. In a very recent work, Saberian et al. \cite{SaGhKi15} found the closed form solution to the problem of optimal fractionation while maintaining multiple simultaneous normal tissue constraints without considering any presumptions about the configuration of the optimal solution. They solved the problem to optimality by instead solving a two-variable linear program with two additional nonlinear constraints.

 An important result emerging from recent work is that the sparing factor of normal tissues and the magnitude of the $\alpha/\beta$ ratio for both normal tissues and the tumor determine the optimal radiation schedule (\cite{badri}, \cite{mizuta2012mathematical}, \cite{SaGhKi15} and \cite{unkelbach2013dependence}). Therefore the optimal fractionation schedule is acutely sensitive to perturbations in these parameters.  One consequence of this sensitivity is the following: an optimal fractionation schedule will have been derived for a fixed set of parameter values (called the \textit{nominal values}), but for a specific patient with a distinctly different set of parameter values this schedule is no longer optimal, and in fact may have poor performance. The uncertainties in radiotherapy treatment can be categorized into two groups: geometric and inter-patient variability. Target volumes take account of geometric uncertainties such as organ motion, inaccuracies or variations in treatment set-up, patient positioning errors and fluctuations in machine output. Several studies addressed these uncertainties using different techniques. Stroom et al \cite{stroom1999inclusion} developed a method for the automatic calculation of planning target volume margins 
as a means for incorporating geometric uncertainties in the region that is irradiated. The traditional approach to dealing with uncertainty in IMRT (considering a margin surrounding the tumor volume)  increases the radiation exposure of healthy tissue. Chan et al. \cite{chan2006robust} developed a robust framework to incorporate uncertainty in the probability distribution that describes breathing motion, and showed that a treatment plan obtained from the robust formulation delivers $38\%$ less dose to the OARs than the traditional solution,
while providing the same level of protection against breathing uncertainty.
In a similar work, Chu et al. \cite{chu2005robust} used a robust optimization approach to find the IMRT treatment plans while considering patient motion and setup uncertainties. More specifically they included uncertain voxel location in their model and as a consequence the delivered dose became a random variable. They designed a mathematical model constructing plans that are more adept at sparing healthy tissue while maintaining the prescribed dose to the target under uncertainty. In \cite{unkelbach2007accounting}, two
methods to account for range uncertainties, one method using a probabilistic approach and the other applying methods from robust linear programming were presented to find optimized treatment plans for intensity modulated proton therapy. Both methods greatly reduced the
sensitivity to range uncertainties of the resulting treatment plans. A modification of
the worst case optimization was applied to a clinical case by Pflugfelder et al. \cite{pflugfelder2008worst}. In addition to the robust optimization, stochastic programming has also been used to account for organ motion and setup errors in IMRT optimization (see \cite{lof1995optimal}, \cite{lof1998adaptive}), e.g. Unkelbach developed a planning method that accounts for the probabilistic
dwelltime of a tumor evaluated from multiple CT scans \cite{unkelbach2004inclusion}. 

Inter-patient variability is due to heterogeneity in patient-specific variables such as the sensitivity of their normal tissues and tumor to radiation, and the growth rate of their tumor. In several cancers there have been multiple subtypes discovered driven by distinct genetic pathways and having distinct phenotypic behaviors such as growth parameters and response to therapy (e.g. glioblastoma \cite{TCGA_GBM}, breast cancer \cite{TCGA_BC}, head and neck cancer \cite{FaWeLi08}, melanoma \cite{MeHaCa14} and many others). A distinct possibility is that there is still significant patient variability within these subtypes. In fact this inter patient heterogeneity is a large reason for the pursuit of personalized medicine \cite{HaCo10}. Given current technologies it is difficult to measure tumor response parameters $\alpha$ and $\beta$ during treatment due to confounding effects such as protracted cell death \cite{FoViAl99}, cell cycle arrest \cite{BeMuMc95}, and radiotherapy mediated immune response \cite{LaErOr12}.
Furthermore, toxicity effects often do not show up until several months or even years after conclusion of therapy, and it is therefore not possible to learn the tumor response properties of normal tissues during treatment. In this paper we concentrate
on the modeling the uncertainties arising in inter-patient variations, which we will use later in an optimization method for radiotherapy scheduling. In a recent unpublished manuscript \cite{Ajdari} (appearing on web after the first version of current manuscript), the authors developed a similar model where the uncertain parameters is assumed to take values in a given interval. Unfortunately this optimization method may lead to an overly pessimistic solution, and furthermore they did not include the uncertainty in the tumor radio-sensitivity parameters ($\alpha$ and $\beta$). To the best of our knowledge, the present study is the first to address the stochastic and linear uncertainty generated by inter patient heterogeneity.

We present a mathematical formulation of the optimal fractionation problem in the presence of multiple normal tissues incorporating uncertainties in model parameters based on the LQ model adjusted for tumor proliferation with a time lag. This formulation allows for the parametric uncertainty to take two forms. First a minimal underlying stochastic model of the uncertain parameters is assumed to be known and every parameter, independently of other entries, takes values in a given interval. We formulate our problem as a model whose solution must be feasible for all realizations of the parameters, and even a small violation of the constraints cannot be tolerated. This method may lead to an overly pessimistic solution, therefore we develop additional models where we assume that the uncertain parameters are characterized by a probability distribution and we reformulate our optimization problem to now insure that our constraints are satisfied with a given probability, and that our objective function achieves a desired level with a given probability. We examine the mathematical properties of the optimal fractionation scheme in various models. The results are discussed in the context of head and neck tumors. As a generalization, we broadly consider the effects of parametric uncertainty on the structure of optimal fractionation schedules.

The organization of the remainder of the paper is as follows. In section two we describe the problem formulation in the setting of fixed parameters (the nominal setting), and then formulate the robust and probabilistic counter parts of this nominal problem for various uncertainty sets. In the next section we describe our solution methods for the problems presented in section two.
In section four we solve our optimization problems for the specific case of head and neck carcinomas and generalize the results to the other tumors. 
In section five we summarize our results and discuss the implications of our findings.

\section{Model of uncertainty and robust formulation}
 
In this section, we first define an objective function derived from the standard linear-quadratic model of radiotherapy response. We next discuss the constraints that are present in our optimization problem, which are derived by maintaining a fixed level of normal tissue damage for a variety of tissue types. Finally we incorporate parameter uncertainty by formulating robust and probabilistic versions of our optimization problem.

\subsection{The nominal formulation}
 
We now consider the problem of finding fractionation schedules that lead to maximal tumor reduction while maintaining acceptable levels of normal tissue damage. The basic linear quadratic (LQ) model states that if a collection of tumor cells are exposed to $N$ fractions of radiation with $d_j$ Gy (SI derived unit of ionizing radiation) in $j^{th}$ fraction, the reproductively viable fraction of cells is given by $e^{-\sum_{j=1}^{N}\alpha d_j+\beta d_j^2}$. However, reproductively viable tumor cells will eventually begin to reproduce, and thus the total surviving fraction of surviving cells is often adjusted to take into consideration the reproduction of tumor cells. A common way to model the repopulation effect is to assume an exponential repopulation process, see e.g., \cite{travis1987isoeffect}.  Thus the net surviving fraction ($S$) due to combined effects of radiation and repopulation after the conclusion of a fractionated radiotherapy treatment is given by
$$
S=e^{-\sum_{j=1}^{N}\alpha d_j+\beta d_j^2}e^{\frac{\ln(2)(T_r-T_k)^+}{T_{e}}}
$$
where $T_r$, $T_e$ and $T_k$ are respectively radiation delivery duration, effective cellular doubling time and kick-off time (or lag before exponential growth begins). The expression $(T_r-T_k)^+$ is defined as $\max(0,T_r-T_k)$. Throughout this paper, we made two important assumptions. First in order to consider the impact of working hour constraints on the objective function, we assume that working hour constraints require that radiation can only be delivered hourly between $8$ am and $8$ pm and five days per week. Second, we assume for every schedule there exist $n$ daily fractions with equal time elapsed between consecutive fractions. If we define $a$ and $r$ as the quotient and remainder of $\frac{N}{n}$, respectively, and $a'$ and $r'$ as the quotient and remainder of $\frac{a}{5}$, respectively,  we can compute $T_r$ as \eqref{Tr} when $a\not=0$. When $a=0$, we simply have $T_r=\frac{8+12\frac{r-1}{n-1}}{24}$
\begin{equation}\label{Tr}
T_r=\begin{cases}7a'+r',& r=0, r'\not=0\\
7(a'-1)+5,& r=0, r'=0\\
7a'+r'+\frac{8+12\frac{r-1}{n-1}}{24},& r\ge1, r'\not=0\\
7(a'-1)+5+\frac{8+12\frac{r-1}{n-1}}{24},& r\ge1, r'=0\end{cases}
\end{equation}

A natural risk associated with radiotherapy is damage to normal tissue near the tumor. A further complication to this toxicity is that in any radiotherapy treatment there are often a large number of normal tissues exposed to radiation. In addition to the existence of a large number of normal parenchymal cells in the clinical target volume
of the respective organ, all tumor volume contains various stromal tissues (e.g. blood vessels and normal connective tissue). In all these normal cells and structures, radiation side-effects may be different (e.g. see the effect on radiation on parallel, serial and dose-volume organs in \cite{joiner2009basic}). A common measure of toxicity for various normal tissues is the biologically equivalent dose or $BED$ \cite{emami1991tolerance}. In particular, assume that for a specific normal tissue of interest the radiation response is characterized by parameters $\alpha_{i}$ and $\beta_{i}$, furthermore assume that this tissue is exposed to $N$ fractions of sizes $\{d_1,\ldots, d_N\}$ respectively, and lastly assume that for fraction $j$ normal tissue is only exposed to $\delta_i d_j$ Gy of radiation for a sparing factor $\delta_i\in (0,1]$.
For each normal tissue we define the maximal toxicity
$$BED_i^{max}=D_i+\delta_i\frac{\beta_i}{\alpha_i}\frac{D_i^2}{N_{i}}$$
where $D_i$ and $N_i$ are tissue specific parameters. Note that $D_i$ is a tissue specific parameter, e.g. it is defined as maximum total dose for the serial normal tissues and maximum mean dose in parallel normal tissues \cite{saberian2014optimal}.
The toxicity constraints for all the OAR is then given by
\begin{equation}\label{BEDoriginal}
\sum_{j=1}^{N}(d_j + \frac{\beta_i}{\alpha_i} \delta_i d_j^2)\leq BED_i^{max}\;\mbox{ for }1\leq i\leq M,
\end{equation}
where $M$ is the number of different normal tissues under consideration.
By taking the natural logarithm of objective function and  using \eqref{BEDoriginal} to model acceptable normal tissue damage, the nominal problem of finding fractionation schedules that lead to maximum tumor damage while maintaining acceptable levels of normal tissue damage can be modeled as

\begin{equation}\label{nominal}
\max_{d_j\ge0,N\in\mathbb{Z}^+} \ \ \sum_{j=1}^{N}\alpha d_j+\beta d_j^2-g(N)
\end{equation}
subject to
$$
\sum_{j=1}^{N}( d_j + \frac{\beta_i}{\alpha_i} \delta_i d_j^2)\le D_i+\delta_i\frac{\beta_i}{\alpha_i}\frac{D_i^2}{N_{i}}, \ \ i=1,\dots,M
$$
where $g(N)=\frac{\ln(2)[T_r-T_k]^+}{T_{e}}$.

\subsection{Modeling uncertainty in radiobiologic parameters}
In order to solve the optimization problem \eqref{nominal} it is vital to know the parameters $\alpha$, $\beta$, $\beta_i/\alpha_i$, and $\delta_i$ since the optimal fractionation schedule will depend on their value (\cite{badri}, \cite{mizuta2012mathematical} and \cite{unkelbach2013dependence}). However, it is quite difficult to obtain accurate measurements of these parameters in a clinical setting and precise estimates of these values are very difficult to find. Furthermore, due to inter-patient heterogeneity it is possible that a wide range of parameter values are possible across the patient population. For example in several cancers there are a multitude of possible mutational pathways responsible for the creation of the tumor, e.g., breast, glioblastoma, and head \& neck. As a result of this situation we are investigate the effect of parametric uncertainty on the solution to problem \eqref{nominal}.

We assume uncertainties presented in LQ model can take two forms: (i) estimation errors for parameters of constant but unknown value, and (ii) stochasticity of random variables. In the first case only the range of the uncertain parameters is known, specifically, we assume parameter $a$ belongs to a symmetric interval $[\bar{a}-l_a,\bar{a} + l_a]$ centered at $\bar{a}$ and for the second scenario we consider $a$ as a continuous random variable with probability density function $f$. In the second case, we are interested in finding the optimized radiotherapy delivery schedule based on two principles: first the nominal values of sensitive parameters are inaccurate and we only know that they lie in a given and second, using the range alone may lead to an excessively high level of conservativeness and the the objective function may suffer as a result.

\subsubsection{Non-probabilistic robust formulation}

As mentioned above the parameters $\alpha$, $\beta$, $\{\delta_i\}_{i=1}^M$ and $\{\beta_i/\alpha_i\}_{i=1}^M$ are subject to uncertainty and may vary amongst patients. For example the values $0.33$Gy and $0.10$Gy are frequently assumed for the ratio $\beta/\alpha$ for late responding normal tissue and tumor tissue  respectively. However these values  should be considered as a rough estimate as there is little evidence \cite{joiner2009basic} to show that these values can be generalized across a wide range of human normal-tissue endpoints and tumor histologies. For the sparing factor $\delta$, there has been a significant amount of effort dedicated to improving the accuracy and precision of radiation therapy delivery in the past decades. However there still exist sources of uncertainty (e.g., patient motion, organ deformation, positioning uncertainty) which make it impossible to achieve full precision in estimating
parameters associated with organ movements in radiotherapy, and thus the exact value of the sparing factor $\delta$ is often not known.

Here our aim is to construct a robust formulation to \eqref{nominal} that is immune to realizations of the uncertain parameters so long as they lie within certain sets. This approach may be the only reasonable alternative when the parameter uncertainty is uniformly distributed, or if no distributional information is available. First we consider the case that the values of parameters  $\alpha$ and $\beta$ are known and we only know that the parameters $\{\beta_i/\alpha_i\}_{i=1}^M$ and $\{\delta_i\}_{i=1}^M$ lie in given intervals. These uncertainty sources can have a detrimental effect on configurations or feasibility of optimal schedules. In this paper, we assume a fixed amount of radiation are delivered to tumor, however the fraction of radiation absorbed by normal tissues is subject to uncertainty. Here we assume for the $i^{th}$ normal tissue, ${\beta_i/\alpha_i}$ and $\delta_i$ are modeled as symmetric and bounded random variables that take lie in given intervals, or equivalently 
\begin{align}\label{uncertaintyset}
\frac{\beta_i}{\alpha_i}&\in [\bar{\frac{\beta_i}{\alpha_i}}-l_{i},\bar{\frac{{\beta_i}}{{\alpha_i}}}+l_{i}], \text{ and } \delta_i\in[\bar{\delta_i}-l_{\delta_i},\bar{\delta_i}+l_{\delta_i}],\ \ i\in\{1,\ldots, M\}.
\end{align}
Formally, the robust counterpart of \eqref{nominal} considering uncertainties defined in \eqref{uncertaintyset} can be written as

\begin{equation}\label{linearrobust1}
\max_{d_j\ge0, N\in\mathbb{Z}^+} \sum_{j=1}^{N}{\alpha} d_j+{\beta} d_j^2-g(N)
\end{equation}
subject to
$$\mathcal{B}_1\cap\mathcal{B}_2\cap\dots\cap\mathcal{B}_{M-1}\cap\mathcal{B}_M$$
where the definition of $\mathcal{B}_1,\dots,\mathcal{B}_M$ and the derivation of the robust counterpart can be found in the appendix.
\subsubsection{Probabilistic optimization models}

Although \eqref{linearrobust1} provides the strongest protection against excessive toxicity in OAR, it is also the most conservative solution and results in less tumor cell kill than achieved by optimizing the nominal formulation. To address this excessive conservativeness, we control the level of flexibility between robustness and performance of the optimal schedule by using a probabilistic formulation that provides a notion of a budget of uncertainty. We view $\alpha$ and $\beta$ as continuous random variables with joint probability density function $f(\cdot,\cdot)$ and we assume that the cdf of $\frac{\beta_i}{\alpha_i}$ and $\delta_i$ in the $i^{th}$ normal tissue are $F_i$ and $G_i$ respectively. In addition we assume that for each $i$ $\frac{\beta_i}{\alpha_i}$ and $\delta_i$ are independent of  all other random variables in the model. We will require that BED in the $i^{th}$ normal tissue does not exceed some level with a high probability. This desire can be naturally expressed by requiring that the BED in the $i^{th}$ normal tissue exceeds the maximum allowable BED, $BED_i^{max}$, with probability at most $1-p_i$, where $p_i$ is some constant close to 1, e.g., 0.95. Furthermore we require that optimized schedules obtained by our robust formulations result in an objective value which exceed level $z$, with probability more than or equal to $p_z$. Computing the above probabilities, we can derive 

\begin{equation}\label{stochrobust1}
\max_{d_j\ge0,N\in\mathbb{Z}^+,z} z-g(N)
\end{equation}
subject to
$$\int_{0}^{\frac{z}{\sum_{j=1}^{N}d_j}}\int_{0}^{\frac{z-\alpha\sum_{j=1}^{N}d_j}{\sum_{j=1}^{N}d_j^2}} f(\alpha,\beta) d\beta d\alpha\le (1-p_z)P(\alpha\ge0,\beta\ge0)$$
$$\mathcal{S}_1\cap\mathcal{S}_2\cap\dots\cap\mathcal{S}_{M-1}\cap\mathcal{S}_M.
$$
Derivation and the definition of  $\mathcal{S}_1,\dots,\mathcal{S}_M$ are shown in the appendix. 
\section{Solution approach}

We now turn our attention to the solution of the optimization problems presented in the previous section. For any fixed $N$, the feasible regions of models described in Section 2 are compact sets and the objective functions are continuous. Therefore by the extreme value theorem of Weierstrass \cite{pierre1969optimization}, optima exist. 

First we use the variable transformation described in \cite{SaGhKi15} to simplify the formulations posed in \eqref{nominal}, \eqref{linearrobust1}, and \eqref{stochrobust1}. The simplification is to introduce the variables
\begin{equation}\label{trsf}
X=\sum_{j=1}^{N}d_j,\ \ Y=\sum_{j=1}^{N}d_j^2.
\end{equation}
We can use the results of \cite{SaGhKi15} and conclude that there is a feasible solution for \eqref{trsf} if and only if $X^2\le NY$ and $X^2\ge Y$. As a consequence, by adding these constraints, we guarantee that the optimal doses $d_i^*$ can be retrieved based on the solution of the adjusted formulations. 

The proliferation term, $g(N)$, does not depend on dose. By including tumor proliferation, we can also optimize the number of treatment sessions $N$. This can be done by first introducing the maximum number of radiation fractions we are willing to administer in the course of radiation therapy, $N_{max}$. The for each $1\leq N\leq N_{max}$ we solve the simplified versions of problems posed in \eqref{nominal} and \eqref{linearrobust1} using transformation introduced in \eqref{trsf} and the two additional constraints $X^2\le NY$ and $X^2\ge Y$. Finally, we choose the $N$ that maximizes the optimal biological effect on the tumor and return the optimal $(X^*,Y^*)$ associated with the optimal $N$ as the global optimal solution.

The feasible region defined by each $\mathcal{B}_i$ or $\mathcal{S}_i$ using the transformations in \eqref{trsf} can be described in the following simplified form
\small
\begin{equation}\label{feasibleregion}
\left\{Y\ge\frac{D_i^2}{N_i}, X+c_iY\le D_i+c_i\frac{D_i^2}{N_i}\right\}\cup\left\{Y\le\frac{D_i^2}{N_i}, X+c_i'Y\le D_i+c_i'\frac{D_i^2}{N_i}\right\},\ \ i=1,\dots,M
\end{equation}
\normalfont
where $c_i$ and $c_i'$ are the coefficients from $\mathcal{B}_i$ or $\mathcal{S}_i$. When $\alpha$ and $\beta$ are fixed, for every $N$, we can replace the conditions $X^2\le NY$ and $X^2\ge Y$ with $Y\ge\theta_NX$ and $Y\le\theta_1X$ (see \cite{SaGhKi15}), respectively, where for $1\leq k\leq N_{max}$, $\theta_i=\frac{Y^{(k)}}{X^{(k)}}$ and the pair $(X^{(k)},Y^{(k)})$ are the coordinates of the intersection $kY=X^2$ with polygon defined by the OAR constraints given by \eqref{feasibleregion}. Therefore in order to solve the optimization problems \eqref{linearrobust1} and \eqref{stochrobust1}  with a fixed value $N\in\{1,\ldots, N_{max}\}$, we need to specify the corners of the convex hull defined by the inequalities in \eqref{feasibleregion}, $Y\ge\theta_NX$ and $Y\le\theta_1X$. In the appendix, Algorithm \ref{alg1} describes how to construct the corners of the polygon sorted in increasing order of their $x$-coordinates $\mathcal{CH}_N=\{(X_1,Y_1),\dots,(X_k,Y_k)\}$. The optimal solution to \eqref{nominal} and \eqref{linearrobust1} for every $N$ occurs at one of these corners. 

For \eqref{stochrobust1} a more specialized approach is required. The new formulation of \eqref{stochrobust1} is
\begin{equation}\label{simplifiedalpha}
\max_{X,Y\ge0,z} z-g(N)
\end{equation}
subject to
\begin{equation}\label{cost}\int_{0}^{\frac{z}{X}}\int_{0}^{\frac{z-\alpha X}{Y}} f(\alpha,\beta) d\beta d\alpha\le (1-p_z)P(\alpha\ge0,\beta\ge0)\end{equation}
\begin{equation}\label{new}\mathcal{S}_1\cap\mathcal{S}_2\cap\dots\cap\mathcal{S}_{M-1}\cap\mathcal{S}_M\end{equation}
\begin{equation}\label{new1}X^2\le N_{max}Y\end{equation}
\begin{equation}\label{new2}X^2\ge Y\end{equation}

In the rest of this section, we will present a method for solving \eqref{simplifiedalpha}-\eqref{new2}. Consider a constrained optimization problem as 
$$\max f(x),\ \ x\in\mathbb{R}^n,$$
$$\text{subject to } g_i(x)\le0, i=\dots,m$$
then we define the $i^{th}$ constraint to be active (at a solution $y$) if $g_i(y)=0$.
We will now show that in optimality, \eqref{cost} is active and the optimal solution always lies on the most restrictive constraint(s), i.e., the constraint(s) that impose the largest restriction on the dose that can be delivered to the tumor.

\begin{lemma}\label{lemmaboudary}
The optimal $X^*$ and $Y^*$ in \eqref{simplifiedalpha} lie on the feasible boundaries of the region defined by \eqref{new}..\eqref{new2} and furthermore constraint \eqref{cost} is active in optimality.
\end{lemma}
We provide the proof of this result in the appendix. Now we discuss the impact of adding $X^2\le N_{max}Y$ and $X^2\ge Y$ on optimal solutions of \eqref{simplifiedalpha}.  
\begin{lemma}\label{yn2}
At the optimal solution to \eqref{simplifiedalpha}, the constraints $X^2\le N_{max}Y$ and $X^2\ge Y$ are either inactive or the optimal solution occurs at corners $(X^{(1)},Y^{(1)})$ and $(X^{(N_{max})},Y^{(N_{max})})$.
\end{lemma}
The proof of this result is provided in the appendix. As a direct result of Lemma \ref{lemmaboudary} and Lemma \ref{yn2}, we know that the optimal pair $(X^*,Y^*)$ lie along the feasible boundary of the convex hull $\mathcal{CH}_{N_{max}}$. For each feasible pair we can find the unique value $\hat{z}(X,Y)$ that gives equality in \eqref{cost}. Thus to solve the optimization problem we find the pair $(X,Y)$ on the feasible boundary that maximizes the function $\hat{z}(X,Y)$.

In order to solve problem \eqref{simplifiedalpha} we can find $\hat{z}_i=\hat{z}(X_i,Y_i)$ for each $1\leq i\leq k$, and then look at $\max\{\hat{z}_i-g(N):1\leq i\leq k\}$. This process will tell us the optimal corner, but it does not necessarily tell us the optimal pair $(X^*,Y^*)$. In order to find the optimal pair it is necessary to consider the edges of the polygon because it is possible to construct numerical examples where the optimal solution does not occur at a corner. 
Note that point $(X_i,Y_i)$ in $\mathcal{CH}_{N_{max}}$ is only connected to the points $(X_{i-1},Y_{i-1})$ and $(X_{i+1},Y_{i+1})$, where $(X_0,Y_0)\equiv (X_k,Y_k)$ and $(X_{k+1},Y_{k+1})\equiv (X_1,Y_1)$.We therefore define the vector valued function for $1\leq i\leq k$
$$
(X_{i}(t),Y_{i}(t))=(tX_i+(1-t)X_{i+1},tY_i+(1-t)Y_{i+1}),\ \ 0\le t \le 1,
$$
and the inverse function

\begin{equation}\label{zi}
z_{i}(t)=\{z-g\left( \lceil X_{i}(t)^2/Y_{i}(t)\rceil\right) |\int_{0}^{\frac{z}{X_{i}(t)}}\int_{0}^{\frac{z-\alpha X_{i}(t)}{Y_{i}(t)}} f(\alpha,\beta) d\beta d\alpha= (1-p_z)P(\alpha\ge0,\beta\ge0)\}.
\end{equation}

Note that the value of $z$ in \eqref{zi} for every pair $(X_{i}(t),Y_{i}(t))$ can be computed using bisection method. The minimum number of treatment sessions for a given $(X,Y)$ is $\lceil X^2/Y \rceil$, thus $g\left( \lceil X_{i}(t)^2/Y_{i}(t)\rceil\right)$ computes the reproduction effect for every $(X_{i}(t),Y_{i}(t))$. We use Algorithm \ref{alg2} in appendix which is designed based on the branch and bound approach to find the optimal solution of \eqref{simplifiedalpha}. On each edge of feasible region, the branching is done on variable $t$ and the global optimal solution is found via searching through all sub-optimal solutions on each edge. The choices for upper and lower bounds in each subproblem is given in the appendix. The optimal number of treatment sessions for an optimal pair of $(X^*,Y^*)$ is $N^*=\lceil (X^*)^2/Y^*\rceil$.

The previous result shows how the solution $(X^*, Y^*, N^*)$ to the simplified versions of problems posed in Section 2 can be found. In \cite{SaGhKi15}, authors proved that $(X,Y)$ transformation is indeed possible and is without loss of optimality. Moreover, we can derive optimal $\{d_1^*,\dots,d_N^*\}$ with the following result.

\begin{theorem}\label{theorem1}
Optimal solution of $\{d_1^*,\dots,d_N^*\}$ retrieved from $(X^*,Y^*,N^*)$ takes one of the following two forms.

\begin{enumerate}
\item $N^*Y^*=(X^*)^2$ for some $N^*\in \{1,\ldots, N_{max}\}$. In this case optimal schedule is given by $d_i^{*}=X^*/N^*$ for $i=1,\ldots, N^*$. Note that if $N^*=1$ then the schedule is hypo-fractionated, and if $N^*>1$ then the schedule is hyper-fractionated.

\item{$(X^*)^2/Y^*$ is not an integer:} In this case, the optimal solution given by the following. Choose a positive integer $j$ less than $(X^*)^2/Y^*$ and set
\begin{equation}\label{semihyper}
d_{1}^{*}=\dots=d_{j}^{*}=\frac{j{X^*}+\sqrt{(N^*-j)(jN^*Y^*-j{X^*}^2)}}{jN^*},\ \ d_{j+1}^{*}=\dots=d_{N^*}^{*}=\frac{{X^*}-jd^*_1}{N^*-j}.\end{equation}

\end{enumerate}
\end{theorem}
The proof of above result can be found in the appendix.

\section{Results}
In this section, we first discuss the effect of uncertainty on the structure of the optimal schedule. Then the application of nominal and robust optimization to the treatment of head and neck tumors via radiotherapy will be discussed. We will describe the data set and parameters that were used in our numerical experiments, then the solution to the nominal and robust optimum dosing schedules will be explored. At the end of this section the sensitivity of the optimal solution to model parameters is studied.
\subsection{Effect of uncertainty on the optimal solution}
\begin{figure}[ht!]
\centering
\includegraphics[width=180mm]{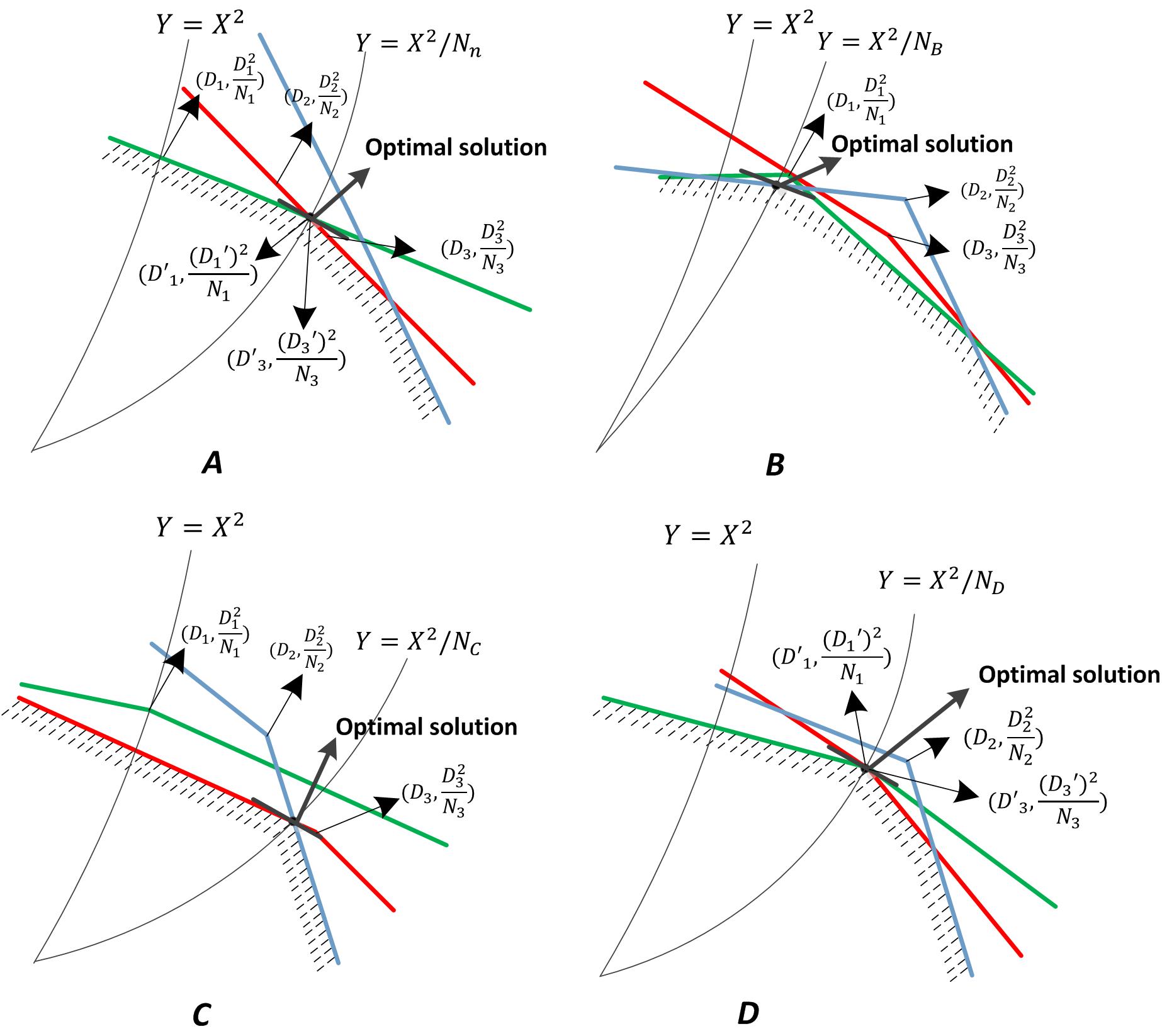}
\caption{This plot shows the effect of uncertainty on the optimal solutions in the presence of three OARs. In each subfigure the feasible region is shaded. In panel (A) we plot the feasible region and optimal solution in the nominal case. Note we also label the points $\left(D_1^\prime,\frac{(D_1^\prime)^2}{N_1}\right)$ and $\left(D_3^\prime,\frac{(D_3^\prime)^2}{N_3}\right)$ as alternative maximum tolerable doses. In panel (B) we consider a scenario where the solution to the stochastic problem results in fewer total fractions, i.e., $N_B<N_n$. In panel (C) we consider different distributions for our uncertain parameters and we have a scenario where the optimal number of fractions in the stochastic problem is greater than the number in the nominal, i.e., $N_C>N_n$. Finally in panel (D) we use the alternative maximum tolerable doses $\left(D_1^\prime,\frac{(D_1^\prime)^2}{N_1}\right)$ and $\left(D_3^\prime,\frac{(D_3^\prime)^2}{N_3}\right)$ and construct a scenario where the optimal number of doses is unchanged by parameter uncertainty, i.e., $N_D=N_n$. If the optimal number of radiation sessions stays the same in the presence of uncertainty, it is required to deliver equal or less doses in the presence of uncertainty ($D$). If we have fewer (more) treatment sessions in probabilistic or robust solution compared to nominal schedule, a reduction in total dose  (total dose squared) will be seen ($B$ and $C$).}
\label{opt}
\end{figure}

Here we study the effects of parametric uncertainty by considering what happens to the optimal solution when the linear and stochastic robust formulations are used instead of the nominal formulations. We use $(X_n,Y_n,N_n)$ and $(X_r,Y_r,N_r)$ to denote the optimal solutions to the nominal and robust (either stochastic or linear) problems. Throughout this subsection, we assume that the nominal and mean values of $\beta_i/\alpha_i$ and $\delta_i$ are equal to  $\bar{\beta_i}/\bar{\alpha_i}$ and $\bar{\delta_i}$, respectively and the probability $p_i$ is greater than $50\%$.  By imposing these two assumptions the feasible region of stochastic and robust problems becomes a subset of the feasible region of nominal problem (see Figure \ref{opt}). Note that the results presented in followings are valid only if we do not allow uncertainty in $\alpha$ and $\beta$ (using same objective function as \eqref{nominal}). We will study the effects of uncertainty in $\alpha$ and $\beta$ in the specific context of head and neck cancer in the next section.

The feasibility region in the  nominal formulation \eqref{nominal} is defined by several inequalities. Note that after using the transformation \eqref{trsf}, these become linear inequalities. By introducing linear or stochastic uncertainty every line segment associated with each inequality will be broken down into two line segments with different slopes where each segment passes through $(D_i,\frac{D_i^2}{N_i})$. There are different possibilities, depending on the amount of uncertainties in model parameters, slope of objective function ($\alpha$ and $\beta$), reproduction rate of the tumor and tumor kick-off time, for how the optimal solution of the robust and stochastic problems relates to the optimal solution of the nominal problem. The feasible region of \eqref{linearrobust1} and \eqref{stochrobust1} can have either more, less or the same number of corner points as compared to the feasible region of \eqref{nominal} (see Figure \ref{opt}). We have three different scenarios for the robust or stochastic optimal schedule.

\begin{enumerate}
\item $N_r<N_n$: In this case we require that the total dose delivered to the tumor decrease, i.e., $X_r<X_n$. However we may have an increase or decrease $Y_r$ depending on the parameters (compare Figure\ref{opt}-A and \ref{opt}-B).
\item $N_r>N_n$: Here we always have $Y_r<Y_n$. We can not say much about $X_r$ and it may be greater or smaller than $X_n$ (compare Figure \ref{opt}-A and \ref{opt}-C).
\item $N_r=N_n$: In this case since the feasible regions of \eqref{linearrobust1} and \eqref{stochrobust1} are subsets of the feasible region of \eqref{nominal}, we require that the total dose and total dose squared delivered to the tumor decrease or stay same. If for two normal tissues, $i$ and $j$, we have $D_i=D_j$, $N_i=N_j$ and $(X_n,Y_n)=(D_i,\frac{D_i^2}{N_i})$, then if the corner $(X_n,Y_n)$ stays feasible after adjusting feasibility region based on model uncertainties, we will have $X_r=X_n$ and $Y_r=Y_n$ (see Figure \ref{opt}-A and \ref{opt}-D).
\end{enumerate} 

To help illuminate the reasoning behind these ideas we give a short proof of (2).

First note that the optimal solution satisfies the condition  $X^2\leq NY$. We know that the objective function is decreasing in $N$, thus the optimal $N$ will be the smallest integer satisfying the constraint  $X^2\leq NY$, i.e. $N_n=\lceil(X^*)^2/Y^*\rceil$. 

Let $m_*=X_*^2/Y_*$ where $(X_*,Y_*)$ are the optimal pair for the nominal problem and note that $m_*$ is not necessarily an integer. Assume $N_r>N_n$ and then obviously  $N_r>N_n\geq m*$. For any real number $m_1\in (N_r-1,N_r]$, let $(X_1,Y_1)$ be the intersection of line $X^2=m_1Y$ with the feasible region of the nominal problem. Since $m_1\geq N_n$ and the slope of line segments constructing this feasible region are negative, then we have $Y_1\leq Y_*$. Next define $(X_2,Y_2)$ as the intersection of the line $X^2=m_1Y$ with the feasible region of the stochastic/robust problem. Since the feasible region of stochastic/robust problem is a subset of the feasible region of the nominal problem defined by line segments with negative slopes, then we have $Y_2\leq Y_1$. Since it is true for every  $m_1\in (N_r-1,N_r]$, then we can set  $m_1=X_r^2/Y_r$, and thus conclude $Y_r\leq Y_1\leq Y_*$. A similar argument shows that if $N_r<N_n$, then we have $X_r<X_*$. 

For model $\eqref{linearrobust1}$, the changes in $X_r$ and $Y_r$ depend on both $N_r$ and the amount of uncertainty in model parameters. Larger uncertainties (large $l_i$) result in larger reductions. In model $\eqref{stochrobust1}$, these reductions not only does depend on $N_r$ and the amount of uncertainty (defined by the variance of uncertain parameters) in the random variables, but also a key factor is the risk tolerance of the decision maker which is defined by $p_i$. If we have $p_i\not=1$, the reduction in $(X_r,Y_r)$ in $\eqref{stochrobust1}$ is smaller than $\eqref{linearrobust1}$.

\subsubsection{Hyper vs Hypo-Fractionation}
One interesting question we can investigate is, when is a hyper-fractionated schedule preferable, and how does this compare to the setting without parameter uncertainty? In order to answer this question we make some simplifying assumptions. First we ignore tumor repopulation, and second we assume that there is tumor radio sensitivity parameters $\alpha$ and $\beta$ are deterministic (i.e., known values). 
If $p_i>0.5$ and $P(\delta_i(\beta_i/\alpha_i)\leq 0)\approx0$, then we will have following different scenarios. 
\begin{itemize}
\item If $(\alpha/\beta)\leq \min_{1\leq i\leq M}{1/b_i}$, then a hypo fractionated schedule is optimal.
\item  If $(\alpha/\beta)\geq \max_{1\leq i\leq M}{1/b_i^\prime}$, then a hyper fractionated schedule with $N_{max}$ fractions is optimal.
\item If $\min_{1\leq i\leq M}{1/b_i}<(\alpha/\beta)<\max_{1\leq i\leq M}{1/b_i^\prime}$, either a hypo or a hyper or an unequal multiple dosage solution can be optimal. 
\end{itemize}
The constants $b_i$ and $b_i^\prime$ are related to the distributions of $\beta_i/\alpha_i$ and $\delta_i$ and are defined in the appendix.

In the nominal setting in the absence of tumor reproduction, \cite{SaGhKi15} showed the following.
\begin{itemize}
\item If $(\alpha/\beta)\leq\min_{1\leq i\leq M}(\bar{\alpha_i}/\bar{\beta_i})/\bar{\delta_i}$, then a hypo fractionated schedule is optimal.
\item If $(\alpha/\beta)\geq\max_{1\leq i\leq M}(\bar{\alpha_i}/\bar{\beta_i})/\bar{\delta_i}$, then a hyper fractionated schedule with $N_{max}$ fractions is optimal. 
\item If $\min_{1\leq i\leq M}(\bar{\alpha_i}/\bar{\beta_i})/\bar{\delta_i}<(\alpha/\beta)<\max_{1\leq i\leq M}(\bar{\alpha_i}/\bar{\beta_i})/\bar{\delta_i}$, either a hypo or a hyper or an unequal multiple dosage solution can be optimal.
\end{itemize}

Based on the previous two sets of results we see that we are using $1/b_i$ or $1/b_i^\prime$ instead of $(\bar{\alpha_i}/\bar{\beta_i})/\bar{\delta_i}$. Since $p_i>0.5$, it means $1/b_i<(\alpha_i/\beta_i)/\delta_i<1/b_i^\prime$, therefore we are making $\min_i\{(\bar{\alpha_i}/\bar{\beta_i})/\bar{\delta_i}\}$ smaller and $\max_i\{(\bar{\alpha_i}/\bar{\beta_i})/\bar{\delta_i}\}$ greater. The result of this is that we increase the chance of having $\min_i(1/b_i)<(\alpha/\beta)<\max_i(1/b_i^\prime)$, where either a hypo or a hyper or an unequal multiple dosage solution can be optimal.

\subsection{Application to head and neck tumors}

In order to estimate head and neck tumor radiobiologic parameters, we use the data set in \cite{rezvani1993sensitivity}. To improve estimation accuracy, trials with the same properties such as total dose administered, number of fractions and treatment duration were merged. Model fit is carried out by minimizing the weighted error between the model predictions of survival probability and the observed values of survival probabilities in trials associated with different schedules. In particular, the results of $K$ trials have been considered. The trial outputs are survival fraction for each trial $f_1,\dots,f_K$. We assume that the tumor cell population regrows exponentially after irradiation with a time lag of $T_k$ and rate $\gamma$. Then following fractionated radiotherapy with $X=\sum_{i=1}^{N}d_i$ and $Y=\sum_{i=1}^{N}d_i^2$, the population of tumor cells $T$ units of time after start of therapy is given by
\begin{equation}\label{estimationwith}
N(T)=N(0)\exp[-\alpha X-\beta Y]\exp[\gamma(T-T_k)^+].
\end{equation}
As a simplification we say that recurrence is only detectable if $N(T)/N(0)\ge 1$, i.e., if the tumor is bigger than its size at the start of therapy. There are $K$ total trials and the total radiation in trial $i$ is $X_i$ and the sum of the doses squared in trial $i$ is $Y_i$, and that the radiotherapy lasted for ${T_r}_i$ days. Then from \eqref{estimationwith} we want to choose the distribution of $\alpha$ and $\beta$ such that for each $1\le i\le K$
$$P(\exp[-\alpha X-\beta Y]\exp[\gamma(T-T_k)^+]\le 1)=f_i$$
Assume that the distributions of $\alpha$ and $\beta$ are characterized by the joint density $f(x; y; \theta)$ where $\theta$ is a parameter that specifies the distribution. We define the probability that $(\alpha,\beta)$ take values in some set as (with the $\theta$ dependence explicit)
$$P_\theta(a_1\le\alpha\le a_2,b_1\le \beta_1\le b_2)=\int_{a_1}^{a_2}\int_{b_1}^{b_2}f(x,y;\theta)dxdy.$$ We assume that $\theta$ takes values in the space $\Theta$. For each trial $1\le i\le K$ we define function 
$$\phi_i(\theta)=P_\theta(\exp[-\alpha X_i-\beta Y_i]\exp[\gamma(T-T_k)]\le 1).$$
Then our procedure for finding the best parameter set is to solve the minimization problem 
$$\min_{\theta\in\Theta}\sum_{i=1}^{K}n_i(\phi_i(\theta)-f_i)^2$$
where $n_i$ is the number of patients in $i^{th}$ trial. Simulated annealing algorithm is utilized to find the optimal values of above model. We assume that the radiosensitivity parameters of LQ model, $\alpha$ and $\beta$ are distributed based on two independent normal distributions with means $\mu_\alpha$ and $\mu_\beta$ and standard deviations $\sigma_\alpha$ and $\sigma_\beta$, respectively. The reproduction rate $\gamma$ ($=\frac{\ln(2)}{T_e}$) and Kick-off time, $T_k$, for head and neck were selected to be $0.003$ per day and $21$ days respectively  \cite{rezvani1993sensitivity}. The parameter $T$ was set to be 5 years and the nominal values of ${\alpha}$ and ${\beta}$ were set to be equal $\mu_\alpha$ and $\mu_\beta$. All parameters are summarized in Table \ref{tumor}.  

We consider six different normal tissues involved in the treatment of head and neck carcinomas (\cite{castadot2010adaptive} and \cite{saibishkumar2007sparing}). The nominal values and confidence intervals for $\beta/\alpha$ for various normal tissues were extracted from \cite{ruifrok1992fractionation}, \cite{turesson1989repair}, \cite{Maciejewski1990oralcavityl}, \cite{meeks2000calculation}, \cite{orton1988unified}, \cite{withers1995late}, \cite{maciejewski1986alpha} and \cite{pan2007physical} and are listed in Table \ref{normal}. We assume that the ratio of $\{\beta_i/\alpha_i\}_{i=1}^{i=M}$ for normal tissues are distributed based on normal distributions with means $\mu_{(\beta/\alpha)_i}$ and standard deviations $\sigma_{(\beta/\alpha)_i}$. The values of means $\mu_{(\beta/\alpha)_i}$ were set to the average of lower bound and upper bound of confidence intervals reported in above references. Also the standard deviation of $\{\beta_i/\alpha_i\}_{i=1}^{i=M}$ associated with different normal tissues were computed based on their confidence intervals given in references.  Normal distributions with parameters $\mu_{\delta_i}$ and $\sigma_{\delta_i}$ are considered for $\{\delta_i\}_{i=1}^{i=M}$. Mandible and spinal cord are considered as serial structures and the data reported in \cite{fiorino2006significant} is utilized to compute their distribution parameters. Brain stem and parotid glands are assumed to be a serial and parallel tissues, respectively, and their parameters are estimated from data reported in \cite{saibishkumar2007sparing}. In these papers, average and standard deviation of dose absorbed by a normal tissue for a given dose radiated to the tumor are reported. In \cite{castadot2010adaptive}, the values for planned dose, actually delivered dose and re-planned dose have been reported. We used these values to obtain a range for sparing factors for larynx and skin. Skin is considered a serial structure and larynx is considered a parallel structure. Nominal values of $\{\beta_i/\alpha_i\}_{i=1}^{i=M}$ and sparing factors are set to the $\mu{(\beta/\alpha)_i}$ and $\mu_{\delta_i}$, respectively. The tolerance dose values for various normal tissues were computed from \cite{emami1991tolerance}, delivered in $35$ fractions. We assume that patients may be treated at most in seven weeks and they visit the clinics three times a day, $n=3$. By considering 5 working days every week, we can compute the maximum number of allowable fractions as $N_{max}=7\times 5\times n=105$. 

In order to understand the effects of parametric uncertainty on the structure of the optimal schedule, we consider one additional setting for $\beta$. In particular we change the values of $\mu_\beta$ and $\sigma_\beta$ to $0.0001$ (we call this scenario case 2 and the original values are case 1). The consideration of small values for $\beta$ enables us to study the effect of uncertainty on schedules with a large number of fractions. The maximum dose constraints of $32\ \ Gy$ for parotid glands results in optimal schedule with small values of total dose. Since in clinical practice larger dose has been used, we report optimal solution for two cases, including parotid glands in our constraint set and excluding it. Table \ref{opttable} displays the optimum schedule for different models (optimal doses can be calculated from Theorem \eqref{theorem1}). 
Our numerical results for head and neck tumor show that the presence of uncertainty changes the optimal schedule to a schedule with larger dose delivered in more fractions (smaller total dose squared) for case 1 and almost same total dose delivered in smaller fractions for case 2.  As expected, the value of the objective function $z^*$ in \eqref{stochrobust1} is decreasing in the probability of having the actual tumor BED less than the optimal value of $z^*$. The $z^*$ value drops from $18.01$ to $6.22$ if we increase $p_z$ from $50\%$ to $90\%$.

\begin{figure}[ht!]
\centering
\includegraphics[width=160mm]{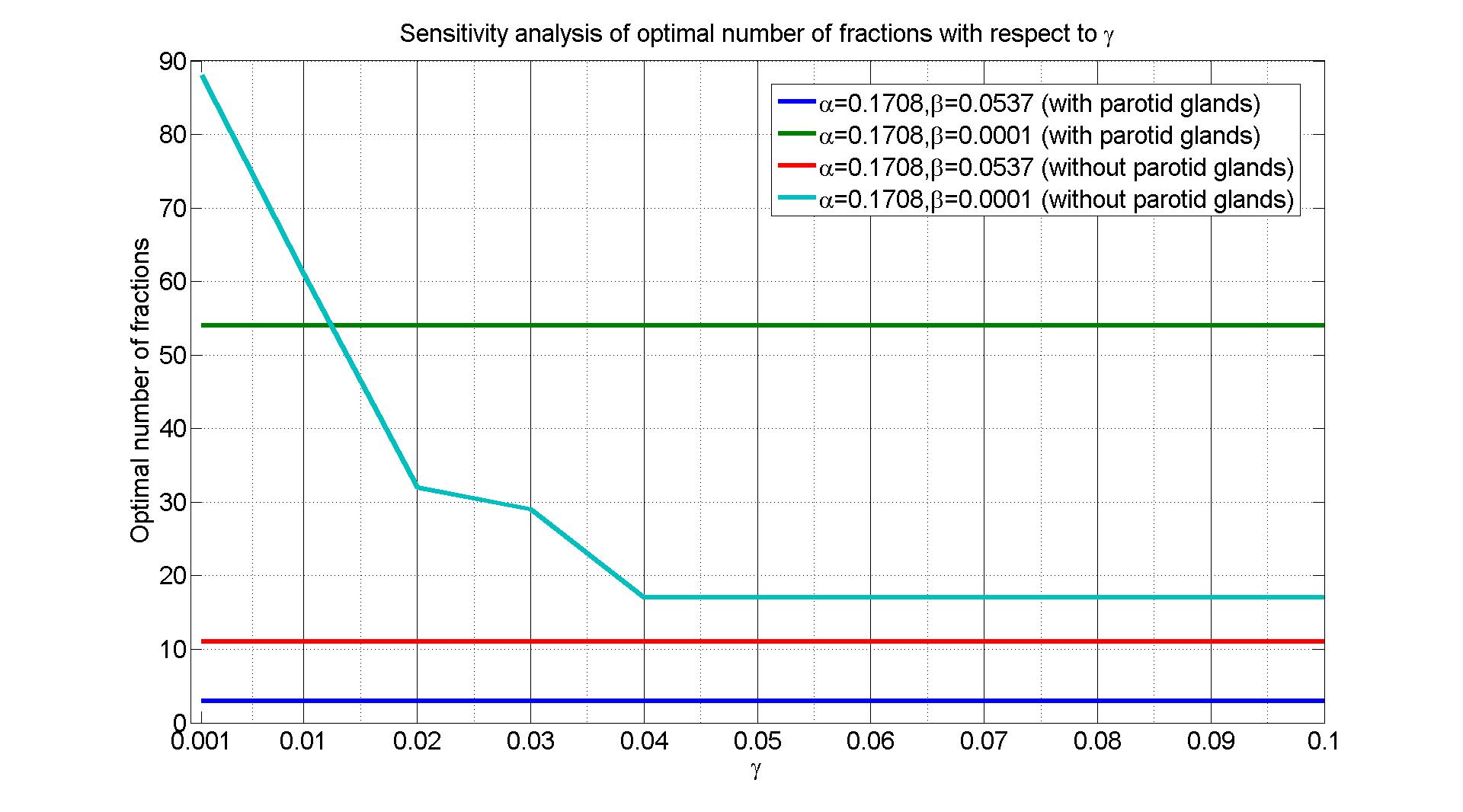}
\caption{This plot shows the sensitivity of treatment session in \eqref{stochrobust1} assuming $p_i=p_z=95\%$ with respect to tumor growth rate $\gamma$ assuming $T_k=7$ days.  For short schedules the objective solution is robust to drifts in $\gamma$ and for fast growing tumors, long treatment time have a negative effect on the treatment outcome. }
\label{gamma}
\end{figure}

Figure \ref{gamma} plots the $N^*$ in \eqref{stochrobust1} for different values of $\gamma=\frac{\ln(2)}{T_e}$. The optimal value of the objective function is a non-increasing function in $\gamma$. For short schedules, since the proliferation effect is negligible  $g(N)\approx 0$, the objective solution is robust to drifts in $\gamma$. However in fast growing tumors, long treatment times have a negative effect on the treatment outcome. 

\section{Conclusion}
In this work, we have analyzed the problem of finding optimal radiation administration schedules considering various types of normal tissues in the presence of model parameter uncertainty. In particular, we aimed to identify the optimized total dose, number of fractions, dose per fraction and treatment duration for a variety of formulations considering different types of uncertainty. We used the traditional linear quadratic model including tumor proliferation to investigate the dynamics of radiation response considering two uncertainty sets. First we assumed that only a range of possible values is known for the model parameters, $\frac{\beta}{\alpha}$ and sparing factors of normal tissue, $\delta$, are known. We presented robust formulations of our optimization problem that are immune to realizations of the uncertain parameters so long as they lie within their respective ranges. Since using the ranges alone may lead to an excessively high level of conservativeness, in the second phase, we adjusted our formulations for the cases that uncertain parameters are distributed as continuous random variables with known probability density functions. Here we imposed the risk aversion factors in the objective function and the feasibility of constraints using some pre-defined probabilities.

We used the transformation introduced in \cite{SaGhKi15}, defining the total radiation as $X$ and sum of doses squared as $Y$, and showed that our problem can be significantly simplified and easily solved in two dimensions when uncertainty in $\alpha$ and $\beta$ are disregarded in the problem. In this case we observed that if the constraint $X^2=NY$ is active in optimality, then the largest possible BED to the tumor can be given in an equal-dosage schedule, otherwise the optimal solution is a semi-equal dosage schedule.  When we consider $\alpha$ and $\beta$ as two continuous random variables, the problem becomes more challenging and optimal solution can happen at a non-corner point. In this case, first we have shown that the optimal value occurs at the boundaries of the feasible region defined by normal tissues BED constraints. We the designed a branch and bound algorithm to solve these stochastic models to optimality. Saberian et al. \cite{SaGhKi15} have recently proposed a method to extract the optimal doses $d_1^*,d_2^*,\dots,d^*_N$ given $X^*$, $Y^*$. However their approach fails if the optimal number of radiation sessions becomes larger than $2$ and $(X^*)^2/Y^*$ is not an integer (note that $d_1$ is not necessarily a positive real number in case $3$ of Theorem 1 in \cite{SaGhKi15} when $(X^*)^2/Y^*>2$). Here we showed that a semi-equal dosage schedule is optimal where $d_1^*=\dots=d_j^*$ and $d_{j+1}^*=\dots=d_{N^*}^*$ for an integer $j<\frac{X^2}{Y}$.   

As a generalization of our results, we observed that when the presence of uncertainty does not change the structure of the optimal solution, it is preferred to administer same or smaller total dose and total dose squared. However if we have larger (smaller) treatment sessions in probabilistic or robust solution compared to nominal schedule, a reduction in total dose squared (total dose) will be seen.

Using data gathered previously \cite{rezvani1993sensitivity}, we parametrized the uncertainty in $\alpha$ and $\beta$ to investigate the behavior of optimal schedules for the head and neck tumors. For the numerical results, we assumed that the head and neck cancer site includes six normal tissues, spinal cord, brain stem, skin, mandible, larynx and parotid glands. The uncertainties in normal tissues have been estimated based on various data sets in the literature. The nominal optimal solution is a hypo-fractionated schedule changing to a schedule with larger total dose delivered in more fractions in the presence of parameter uncertainty. We found that when we consider small values of $\beta$, the optimal schedule is a hyper-fractionated schedule with maximum allowable fractions. In this case the robust solution has an insignificant change in the optimal total dose and total dose squared for different schedules, however the optimal number of fractions decreases in some cases. We saw that as the tumor regrowth rate increases, shorter treatment are preferable. 

There are several possible extensions to this work that we plan to consider in the future. For example, this work does not incorporate spatial structure of the tumor, including possible spatial heterogeneities in the parameters $\alpha$ and $\beta$.
Another possible extension is the incorporation of repair effects, this would be useful if we wanted to consider shorter inter fraction periods. Lastly, it would be interesting to incorporate immune response and how inter-patient heterogeneity in immune response could impact the design of optimal fractionation schedules (see \cite{LaErOr12}).
\clearpage
\section{Appendices}

\subsection{Derivation of the robust reformulations}

\subsubsection{Non-probabilistic robust formulation}
In this section, we derive a computationally tractable solution to the robust optimization
problem \eqref{nominal}. In the non-probabilistic robust formulation we do not allow any violation of the normal tissue constraints for any parameters taking values in the sets \eqref{uncertaintyset}. Therefore the robust counter part of \eqref{nominal} associated with uncertainty sets defined in \eqref{uncertaintyset} is found by solving

\begin{equation}\label{p1}
\max_{d_j\ge0,N\in\mathbb{Z}^+} \ \ \sum_{j=1}^{N}{\alpha} d_j+{\beta} d_j^2-g(N)
\end{equation}
subject to
\begin{align*}
&\sup\left\{\left( \sum_{j=1}^{N}d_j^2-\frac{D_i^2}{N_{i}}\right) \delta_i\frac{\beta_i}{\alpha_i}\Big| {\frac{\beta_i}{\alpha_i}}\in[\bar{\frac{\beta_i}{\alpha_i}}-l_{i},\frac{\bar{\beta_i}}{{\alpha_i}}+l_{i}] \text{ and } \delta_i\in[\bar{\delta_i}-l_{\delta_i},\bar{\delta_i}+l_{\delta_i}]\right\}\le D_i-\sum_{j=1}^{N}d_j  \ \ \forall i.\\
\end{align*}
Note that when  $\sum_{j=1}^{N}d_j^2\ge \frac{D_i^2}{N_{i}}$, the supremum happens when $\frac{\beta_i}{\alpha_i}$ and $\delta_i$ take their upper bounds in the sets \eqref{uncertaintyset}, otherwise the supremum is achieved in lower bounds of $\frac{\beta_i}{\alpha_i}$ and $\delta_i$ defined in \eqref{uncertaintyset}. We now replace the problem \eqref{p1} by a formulation using the supremum of BED constraints:
\begin{equation}\label{p2}
\max_{d_j\ge0, N\in\mathbb{Z}^+} \ \ \sum_{j=1}^{N}{\alpha} d_j+{\beta} d_j^2-g(N)
\end{equation}
subject to
$$\mathcal{B}_1\cap\mathcal{B}_2\cap\dots\cap\mathcal{B}_{M-1}\cap\mathcal{B}_M$$
where $\mathcal{B}_i$ is defined as follows
$$\mathcal{B}_i=\left\{\sum_{j=1}^{N}d_j^2\ge \frac{D_i^2}{N_{i}}, \sum_{j=1}^{N}d_j+a_i\sum_{j=1}^{N}d_j^2\le D_i+a_i\frac{D_i^2}{N_{i}}\right\}\cup\left\{\sum_{j=1}^{N}d_j^2\le \frac{D_i^2}{N_{i}}, \sum_{j=1}^{N}d_j+a_i'\sum_{j=1}^{N}d_j^2\le D_i+a_i'\frac{D_i^2}{N_{i}}\right\}$$
and $a_i=(\bar{\delta_i}+l_{\delta_i})(\bar{\frac{\beta_i}{\alpha_i}}+l_i)$ and $a_i'=(\bar{\delta_i}-l_{\delta_i})(\bar{\frac{\beta_i}{\alpha_i}}-l_i)$.

\subsubsection{Probabilistic optimization models}
We now describe a formulation that assumes that $\alpha$, $\beta$, and $\frac{\beta_i}{\alpha_i}$ and $\delta_i$ are random variables with known probability distributions, and we use our knowledge of this uncertainty to enforce the constraints in a probabilistic fashion. We require that the probability of violation of the BED constraint in $i^{th}$ OAR is at most $1-p_i$. Written mathematically we have
\begin{equation}\label{pr1}
P\left(\sum_{j=1}^{N}(d_j+\delta_i\frac{\beta_i}{\alpha_i}d_j^2)\le D_i+\delta_i\frac{\beta_i}{\alpha_i}\frac{D_i^2}{N_{i}} \Big |{\frac{\beta_i}{\alpha_i}}\ge 0,\delta_i\ge0\right)\ge p_i\ \ \forall i.
\end{equation}
Note that from a biological point of view, it is impossible for $\alpha$, $\beta$, $\frac{\beta_i}{\alpha_i}$ and $\delta_i$ to take negative values, and since in literature it is often assumed that these parameters are normally distributed, we need to add the non-negatively conditions above. By knowing the cdf of $\frac{\beta_i}{\alpha_i}$ and $\delta_i$ (recall that we assume these random variables are independent), we can easily compute the cdf of their product. Let $H_i$ be the cdf of $\frac{\beta_i}{\alpha_i}\delta_i$. In order to satisfy \eqref{pr1} for random variables $\delta_i$ and $\frac{\beta_i}{\alpha_i}$ we must have:
\begin{equation}\label{pr11}
P\left((\sum_{j=1}^{N}d_j^2-\frac{D_i^2}{N_{i}})\delta_i\frac{\beta_i}{\alpha_i}\le D_i-\sum_{j=1}^{N}d_j,\delta_i\frac{\beta_i}{\alpha_i}\geq 0 \right)\ge p_i\bar{H_i}(0)\ \ \forall i.
\end{equation}
Note that we assume $\delta_i\in [0,1]$ and thus the event $\{\frac{\beta_i}{\alpha_i}\ge 0,\delta_i\ge0\}$ is equivalent to $\{\delta_i\frac{\beta_i}{\alpha_i}\geq 0\}$.
If we consider the events in \eqref{pr11} for the positive and negative values of $\sum_{j=1}^{N}d_j^2-\frac{D_i^2}{N_{i}}$, we can rewrite the condition \eqref{pr11} as $\cap_{i=1}^M\mathcal{S}_i$ where
 \small
 $$\mathcal{S}_i=\left\{\sum_{j=1}^{N}d_j^2\ge \frac{D_i^2}{N_{i}}, \sum_{j=1}^{N}d_j+b_i\sum_{j=1}^{N}d_j^2\le D_i+b_i\frac{D_i^2}{N_{i}}\right\}\cup\left\{\sum_{j=1}^{N}d_j^2\le \frac{D_i^2}{N_{i}}, \sum_{j=1}^{N}d_j+b_i'\sum_{j=1}^{N}d_j^2\le D_i+b_i'\frac{D_i^2}{N_{i}}\right\}$$
 \normalsize
and $b_i=H_i^{-1}(1-(1-p_i)\bar{H}_i(0))$ and $b_i'=H_i^{-1}(1-p_i\bar{H}_i(0))$.

 We require that optimized schedules obtained by our robust formulation result in an objective value which exceeds level $z$, with a probability at least $p_z$, i.e.,
\begin{equation}\label{pr2}
P\left(\sum_{j=1}^{N}({\alpha} d_j+{\beta} d_j^2)\ge z\Big |{\alpha}\ge 0,{\beta}\ge 0\right)\ge p_z.
\end{equation}
Note that for a fixed $N$, $g(N)$ is a constant and does not depend on $d_i$ and therefore we can remove it from above probability. We can simplify \eqref{pr2} to get the constraint
$$
\int_{0}^{\frac{z}{\sum_{j=1}^{N}d_j}}\int_{0}^{\frac{z-\alpha\sum_{j=1}^{N}d_j}{\sum_{j=1}^{N}d_j^2}} f(\alpha,\beta) d\beta d\alpha\le (1-p_z)P(\alpha\ge0,\beta\ge0).
$$

\subsection{Proof of technical Lemma \ref{lemmaboudary}}
\begin{proof}
Assume $z^*$,$X^*$ and $Y^*$ are optimal points to \eqref{simplifiedalpha}. If $X^*$ and $Y^*$ lie in the interior of the feasible region, then there exist $\Delta X>0$ and $\Delta Y>0$ such that the pair $(X',Y')$, where $X'=X^*+\Delta X$ and $Y'=Y^*+\Delta Y$, is a feasible solution. The left hand side of \eqref{cost} is a decreasing function in $X$ and $Y$, therefore we can replace $(X^*,Y^*)$ with $(X',Y')$ and increase $z$ without violating feasibility of constraints set defined in \eqref{cost} ... \eqref{new2}. Therefore there exists a feasible $z$ which is strictly greater than $z^*$ and it contradicts the assumption that $z^*$ is an optimal solution to our problem. Therefore the optima must lie on the boundaries of the feasible region. Also if \eqref{cost} is not active in optimality, we have

$$\int_{0}^{\frac{z}{X}}\int_{0}^{\frac{z-\alpha X}{Y}} f(\alpha,\beta) d\beta d\alpha< (1-p_z)P(\alpha\ge0,\beta\ge0)$$
and we can increase $z$ without leaving feasible region which contradicts the optimality assumption. 
\end{proof}
\subsection{Proof of technical Lemma \ref{yn2}}
\begin{proof}
Let $\mathcal{CH}_{N_{max}}$ be the feasible region of \eqref{simplifiedalpha} defined by \eqref{new}, \eqref{new1} and \eqref{new2}. Based on lemma \ref{lemmaboudary}, the optimal solution of \eqref{simplifiedalpha} lies on the feasible boundaries of $\mathcal{C}_{N_{max}}$. If $X^2\le N_{max}Y$ and $X^2\ge Y$ are redundant constraints ($(X^*,Y^*)$ obtained by ignoring these constraints satisfy these constraints), then $X^2\le N_{max}Y$ and $X^2\ge Y$ are inactive constraints in optimality. Otherwise, consider the three corners $p_1=(0,0)$, $p_2=(X^{(1)},Y^{(1)})$ and $p_3=(X^{(N_{max})},Y^{(N_{max})})$. As we move from $p_1$ toward $p_2$ or from $p_1$ toward $p_3$, we can increase both $X$ and $Y$. At the end of two line segments $\overrightarrow{p_0p_1}$ and $\overrightarrow{p_0p_2}$, we are at a feasible solution ($p_2$ or $p_3$) with maximum $X$ and $Y$, and since the objective function in \eqref{simplifiedalpha} is increasing in both $X$ and $Y$, we see that the maximal value of $z$ is obtained at either $p_1$ or $p_2$.
\end{proof}
\subsection{Proof of Theorem \ref{theorem1}}
\begin{proof}
To establish the result in case 1, we can easily check that equation \eqref{trsf} holds for $d_1,\ldots, d_{N^*} =X^*/N^*$ if $N^*Y^*=(X^*)^2$. In the second scenario, from straightforward calculations, we observe that there is always a solution to our problem in the following form:
$$
d_{1}^{*}=\dots=d_{j}^{*}=d,\ \ d_{j+1}^{*}=\dots=d_{N^*}^{*}=w.
$$
We can now solve for $w$ and $d$ in \eqref{trsf}, and establish that 
$$
w=\frac{X^*-jd}{N^*-j}
$$
and
$$
d=\frac{jX^*+\sqrt{j^2(X^*)^2-jN^*((X^*)^2-(N^*-j)Y^*)}}{jN^*}=\frac{jX^*+\sqrt{(N^*-j)(jN^*Y^*-j(X^*)^2)}}{jN^*}.
$$
It then remains to establish that $d$ and $w$ are non-negative real numbers. First observe that  we require that $j\leq(X^*)^2/Y^*\leq N^*$. It follows from this that $d$ is a positive real number, and thus $w$ is a real number as well. It then remains to establish that $w$ is non-negative. This is of course equivalent to showing that $X^*-jd>0$. Note that

\begin{align*}
X^*-jd&=\frac{1}{N^*}\left[(N^*-j)X^*-\sqrt{j(N^*-j)(N^*Y^*-(X^*)^2)}\right]\\
&=
\frac{1}{N^*}\left[(N^*-j)X^*-\sqrt{jY^*(N^*-j)(N^*-(X^*)^2/Y^*)}\right]
\end{align*}
and therefore
\begin{align*}
X^*-jd>0&\Leftrightarrow \frac{(X^*)^2}{Y^*}>j.
\end{align*}
The result then follows from our conditions on the integer $j$.
\end{proof}
\clearpage
\subsection{Algorithms}
\subsubsection{A Feasible Region Creator Algorithm}
\begin{algorithm}
\caption{Constructing the corners of feasible region defined by \eqref{feasibleregion}, $X^2\ge Y$ and $X^2\le NY$}
\begin{algorithmic}[1]\label{alg1}
  \scriptsize
  \STATE Define $\mathcal{L}_1$ as the set of all line segment connecting $(0,\frac{D_i}{c_i}+\frac{D_i^2}{N_i})$ and $(D_i,\frac{D_i^2}{N_i})$ and $\mathcal{L}_2$ as the set of all line segment passing through $(D_i,\frac{D_i^2}{N_i})$ and $(D_i+c_i'\frac{D_i^2}{N_i},0)$ for $i=1,\dots,M$.
  \STATE Use Bentley-Ottmann algorithm \cite{bentley1979algorithms} for listing all crossings in the set of $\{\mathcal{L}_1\cup\mathcal{L}_2\}$, call it $\mathcal{L}$.
  \STATE Let $\mathcal{L}$ be $\mathcal{L}\cup\{(0,\frac{D_{i_1}}{c_{i_1}}+\frac{D_{i_1}^2}{N_{i_1}})\}\cup\{(D_{i_2}+c_{i_2}'\frac{D_{i_2}^2}{N_{i_2}},0)\}$ where $i_1=argmin_k\{\frac{D_{k}}{c_{k}}+\frac{D_{k}^2}{N_{k}}\}$ and $i_2=argmin_k\{D_{k}+c_{k}'\frac{D_{k}^2}{N_{k}}\}$
  \STATE Compute $\mathcal{V}=\{(v_1,v_1'),\dots,(v_p,v_p')\}$, as the set of all pairs in $\mathcal{L}$ satisfying \eqref{feasibleregion} for all $i=1,\dots,M$, sorted in increasing order by the $x$-coordinate.
  \STATE $CH_1\leftarrow\{(0,0)\}$
  \STATE For $i\leftarrow1$ to $p-1$
  \STATE $\mbox{ }$ Compute $(x_i,y_i)$ as the intersection of line passing through $(v_i,v_i')$ and $(v_{i+1},v_{i+1}')$ and $y=x^2$.
  \STATE $\mbox{ }$ If $v_i\le x_i< v_{i+1}$ and $y_i>0$
  \STATE $\mbox{ }\quad$index[1]$\leftarrow i$
  \STATE $\mbox{ }\quad$ $CH_1\leftarrow CH_1\cup\{(x_i,y_i)\}$
  \STATE $\mbox{ }\quad$Break
  \STATE $\mbox{ }$ End If
  \STATE End For
  \STATE For $i\leftarrow2$ to $N_{max}$
  \STATE $\mbox{ }$ $CH_i\leftarrow CH_{i-1}$ 
  \STATE $\mbox{ }$ For $j\leftarrow\text{index}[i-1]$ to $p$ 
  \STATE $\mbox{ }\quad$ If $v_j^2\le iv_j'$
  \STATE $\mbox{ }\quad\quad$ $CH_i\leftarrow CH_{i}\cup \{(v_j,v_j')\}$ 
  \STATE $\mbox{ }\quad\quad$ index$[i]$ $\leftarrow j$ 
  \STATE $\mbox{ }\quad$ Else 
  \STATE $\mbox{ }\quad\quad$ if $j==$index$[i-1]$ then index$[i]$ $\leftarrow$index$[i-1]$, else index$[i]$ $\leftarrow j-1$ 
  \STATE $\mbox{ }\quad\quad$ Break
  \STATE $\mbox{ }\quad$ End If 
  \STATE $\mbox{ }$ End For 
  \STATE End For 
  \STATE For $i\leftarrow2$ to $N_{max}$
  \STATE $\mbox{ }$ Let $(x_i,y_i)$ be the intersection of $x^2=iy$ and line passing through $(v_{\text{index}[i]},v_{\text{index}[i]}')$ and $(v_{\text{index}[i]+1},v_{\text{index}[i]+1}')$
  \STATE $\mbox{ }$ $CH_i\leftarrow CH_{i}\cup{(x_i,y_i)}$ 
  \STATE End For
\end{algorithmic}
\end{algorithm}

\subsubsection{Branch and bound algorithm}

\begin{algorithm}
\caption{Branch and Bound algorithm for maximization of \eqref{simplifiedalpha}}
\begin{algorithmic}[1]\label{alg2}
  \scriptsize
  \STATE For each $(X_i,Y_i)\in \mathcal{CH}_{N_{max}}$, compute ${z}_i(1)$ and set $z^*=z_j(1)$ and $(X^*,Y^*)=(X_j, Y_j)$ where $j=\argmax_i {z}_i$
  \STATE For $l=1$ to $|\mathcal{CH}_{N_{max}}|-1$ Do:
  \STATE $\mbox{ }\quad$ $i\leftarrow1$.
  \STATE $\mbox{ }\quad$ $lb_t(i)\leftarrow0$, $ub_t(i)\leftarrow1$.
  \STATE $\mbox{ }\quad$ $lb_z(i)=\max\{z(X_l(0),Y_l(0)),z(X_l(1),Y_l(1))\}$, $ub_z(i)=z(\max\{X_l(0),X_l(1)\},\max\{Y_l(0),Y_l(1)\})$.
  \STATE $\mbox{ }\quad$ $a(i)\leftarrow 1$.
  \STATE $\mbox{ }\quad$ While $\sum_{i} a(i)>0$ Do:
  \STATE $\mbox{ }\quad\quad$ ind=find($i|a(i)>0$).
  \STATE $\mbox{ }\quad\quad$ For j=1:length(ind) Do:
  \STATE $\mbox{ }\quad\quad\quad$ $i \leftarrow i+1$.
  \STATE $\mbox{ }\quad\quad\quad$ $lb_t(i)=(lb_t(\text{ind}(j))+ub_t(\text{ind}(j)))/2$, $ub_t(i)=ub_t(\text{ind}(j))$.
  \STATE $\mbox{ }\quad\quad\quad$ Compute
  $$lb_z(i)=\max\{z(X_l(lb_t(i)),Y_l(lb_t(i))),z(X_l(ub_t(i)),Y_l(ub_t(i)))\},\ \ ub_z(i)= z(\max\{X_l(lb_t(i)),X_l(ub_t(i))\},\max\{Y_l(lb_t(i)),Y_l(ub_t(i))\}).$$
  \STATE $\mbox{ }\quad\quad\quad$ If ($ub_z(i)-lb_z(i)>\epsilon$), then $a(i)\leftarrow1$, else $a(i)\leftarrow0$.
  \STATE $\mbox{ }\quad\quad\quad$ $lb_t(\text{ind}(j))= lb_t(\text{ind}(j))$, $ub_t(\text{ind}(j))=(lb_t(\text{ind}(j))+ub_t(\text{ind}(j)))/2$.
  \STATE $\mbox{ }\quad\quad\quad$ Repeat steps 13 and 14 with $\text{ind}(j)$ instead of $i$.
  \STATE $\mbox{ }\quad\quad\quad$ Update $z^*$ and $(X^*,Y^*)$ for any $lb_z(j)>z^*$.
  \STATE $\mbox{ }\quad\quad$ End For Loop.
  \STATE $\mbox{ }\quad\quad$ For every $j$ such that $ub_z(j)<z^*$, $a(j)\leftarrow 0$.
  \STATE $\mbox{ }\quad\quad$ Sort $[lb_t(:),ub_t(:),lb_z(:),ub_z(:),a(:)]^T$ column-wise based on $lb_t(i)$.
  \STATE $\mbox{ }\quad$ End While Loop.
  \STATE End For Loop.
\end{algorithmic}
\end{algorithm}
Note that for every line segment defined by two points $(X_1,Y_1)$ and $(X_2,Y_2)$, we can compute the lower bound using  $$\Phi_{lb}=\max\{z_{1}(0),z_{1}(1)\}$$ and compute the upper bound using
\small $$\Phi_{ub}=\{z-g(\lceil\frac{\min(X_1,X_2)}{\max(Y_1,Y_2)}\rceil\Big|\int_{0}^{\frac{z}{\max(X_1,X_2)}}\int_{0}^{\frac{z-\alpha \max(X_1,X_2)}{\max(Y_1,Y_2)}} f(\alpha,\beta) d\beta d\alpha= (1-p_z)P(\alpha\ge0,\beta\ge0)\}.$$\normalsize On each edge, the branching variable is $t\in[0,1]$. At node $i$, we store the lower bound and upper bound of $t$ as $lb_t(i)$ and $ub_t(i)$. Similarly lower and upper bounds of optimal solution at node $i$ are stored in $lb_z(i)$ and $ub_z(i)$. Note that $a(i)=\{0,1\}$ indicates the state of the node in our optimization tree, if $a(i)=1$ then our node is considered as active node and further partitioning can be proceeded through that branch, otherwise we consider that node as an inactive node. Branching in node $i$ continues in this manner until there are no active nodes in that branch or $ub_z(i)-ub_z(i)<\epsilon$, where $\epsilon$ is the given accuracy for optimal value. Since the objective value of an optimal solution cannot be smaller than a lower bound, active nodes with upper bounds smaller than an existing lower bound can be safely deleted (step 19).

\emph{Remark 1}. \emph{Algorithm \ref{alg2} converges and terminates with certificate proving $\epsilon$-suboptimality.}

Number of line segments in partition $\mathfrak{L}_k$ is $k$. Note that total length of these line segments is $L(\mathfrak{Q}_{\text{initial}})$, so
$$\min_{\mathfrak{Q}\in \mathfrak{L}_k} L(\mathfrak{Q})\le \frac{L(\mathfrak{Q}_{\text{initial}})}{k}$$
and hence for big $k$, at least one line segment has small length and having small length will imply that $ub_z(k)-lb_z(k)$ is small.
\clearpage
\subsection{Figures}

\begin{figure}[ht!]
\centering
\includegraphics[width=180mm]{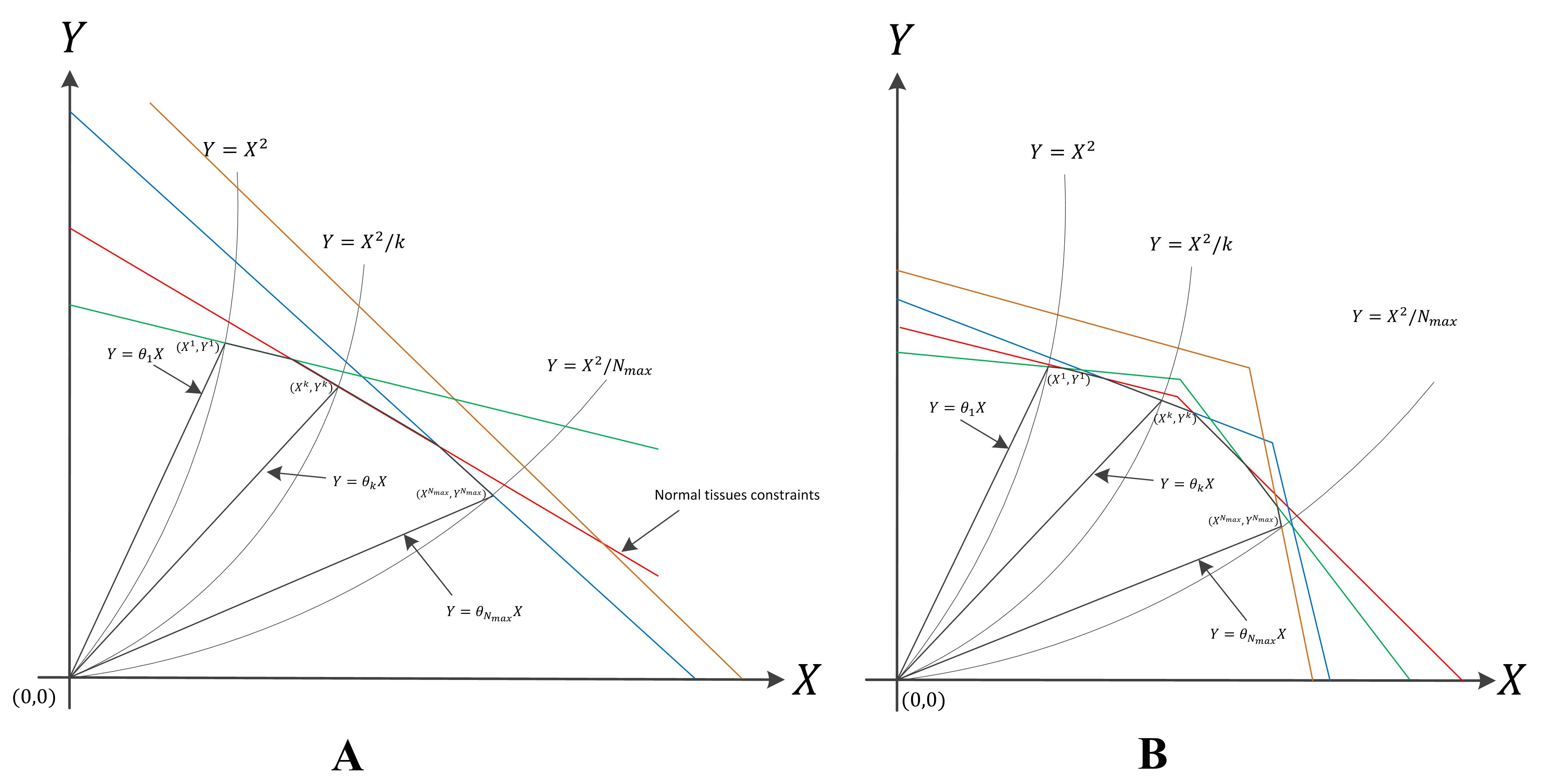}
\caption{Feasible region of optimization problems posed in section 2. A) This plot shows the feasible region of model \eqref{nominal}. For each $N$, the optimal solution occurs at one of the corners of feasible region. B) This plot shows the effect of uncertainty in model parameters on feasible region. Every line segment is broken down into two line segments with different slopes and passing through $(D_i,\frac{D_i^2}{N_i})$. If $\alpha$ and $\beta$ are fixed, the optimal solution for every fixed $N$ lie on one of the corners of $\mathcal{CH}_{N}$. Having $\alpha$ and $\beta$ as random variables, the pair $(X^*,Y^*)$ can be located by searching on all line segments connecting $(X^1,Y^1)$ and $(X^{N_{max}},Y^{N_{max}})$.}
\label{feasible}
\end{figure}

\clearpage
\subsection{Tables}
\begin{table}[ht!]
	\begin{center}
    	\begin{tabular}{ | l | l | l | l | l |}
    	\hline
    	Parameters & Values  & unit \\ \hline
	    $\mu_{\alpha}$ & $0.1708$ &  $1/Gy$ \\ \hline
	    $\mu_{\beta}$  & $0.0537$ &  $1/Gy^2$ \\ \hline
	    $\sigma_{\alpha}$  & $0.2142$ &  $1/Gy$  \\ \hline
	    $\sigma_{\beta}$  & $0.0812$ & $1/Gy^2$  \\ \hline
    	\end{tabular}
    	\caption {Head and neck tumor parameters used for finding optimal schedule}
    	\label{tumor}
	\end{center}
\end{table}
\begin{table}[ht!]
\small
	\begin{center}
    	\begin{tabular}{ | l | | l | l | l | l | l | l | l | }
    	\hline
		 Parameters & Spinal Cord & Brain Stem & Skin  &  Unit \\ \hline
		$\mu_{(\beta/\alpha)}$  & 0.48 & 0.39 & 0.12 &  $1/Gy$ \\ \hline
		$\sigma_{(\beta/\alpha)}$  & 0.09 & 0.05 & 0.01 &  $1/Gy$ \\ \hline
		$(\beta/\alpha-l,\beta/\alpha+l)$  & (0.30,0.67) & (0.30,0.48) & (0.09,0.14) &  $1/Gy$ \\ \hline
		$\mu_{\delta}$& 58.52\% & 74.92\% & 25.29\% &    \\ \hline 
		$\sigma_{\delta}$  & 2.78\% & 3.45\% & 0.75\% &  \\ \hline
		$D_{i}$  & 47 Gy & 50 Gy & 55 Gy & $Gy$ \\ \hline \hline
		Parameters & Mandible & Larynx  & Parotid glands & unit \\ \hline
		$\mu_{(\beta/\alpha)}$   & 0.46 & 0.66 & 0.24 & $1/Gy$ \\ \hline
		$\sigma_{(\beta/\alpha)}$  & 0.05 & 0.30 &  0.07 & $1/Gy$ \\ \hline
		$(\beta/\alpha-l,\beta/\alpha+l)$ & (0.36,0.56) & (0.07,1.25) & (0.10,0.38) & $1/Gy$ \\ \hline
		$\mu_{\delta}$ & 67.59\% &  99.12\% & 40.45\% & \\ \hline
		$\sigma_{\delta}$ & 10.56\% & 0.07\% & 0.97\% & \\ \hline
		$D_{i}$  & 60 Gy & 70 Gy & 32 Gy & $Gy$ \\ \hline
    	\end{tabular}
    	\caption {Normal tissue parameters}
    	\label{normal}
	\end{center}
\end{table}
\begin{table}[ht!]
	\begin{center}
    	\begin{tabular}{ | l | l | l | l | l | l | l | l | l | l |}
    	\hline
    	& & \multicolumn{4}{|c|}{With parotid glands} & \multicolumn{4}{|c|}{Without parotid glands} \\ \hline
    	Parameters & Formulation & $N^*$  & $X^*$ & $Y^*$ & Tumor BED& $N^*$  & $X^*$ & $Y^*$ & Tumor BED \\ \hline
    	\multirow{3}{*}{Case 1} & Nominal & $1$ &  $13.5$ & $182.36$ & $12.10$ & $1$ &  $13.5$ & $182.36$ & $12.10$ \\ 
	    & Robust & $2$ &  $14.26$ & $139.51$ & $9.93$ & $35$ &  $47.00$ & $63.11$ & $11.42$ \\ 
	    & Stochastic & $3$ &  $18.34$ & $124.23$ & $4.36$ & $11$ &  $31.86$ & $95.94$ & $5.08$ \\ \hline
	    \multirow{3}{*}{Case 2} & Nominal & $105$ &  $33.79$ & $10.87$ & $5.69$& $105$ &  $56.26$ & $30.15$ & $9.53$ \\ 
	    & Robust & $62$ &  $32.47$ & $17.01$ & $5.53$& $105$ &  $52.82$ & $26.57$ & $8.95$ \\ 
	    & Stochastic & $54$ &  $32.48$ & $19.80$ & $0.90$& $92$ &  $52.84$ & $30.59$ & $1.41$ \\ \hline
    	\end{tabular}
    	\caption {Optimal solution to problems \eqref{nominal}, \eqref{linearrobust1} and \eqref{stochrobust1} assuming $p_i=p_z=95\%$. In case 1 we assume $\mu_\alpha=0.1708$, $\mu_\beta=0.0537$, $\sigma_\alpha=0.2142$ and $\sigma_\beta=0.0812$ and for case 2 we have  $\mu_\alpha=0.1708$, $\mu_\beta=0.0001$, $\sigma_\alpha=0.2142$ and $\sigma_\beta=0.0001$. }
    	\label{opttable}
	\end{center}
\end{table}

\clearpage
 \bibliographystyle{plain}
 \bibliography{Refs}

\begin{thebibliography}{10}

\bibitem{Ajdari}
A.~Ajdari and A.~Ghate.
\newblock Robust fractionation in radiotherapy.
\newblock {\em available online at
  http://faculty.washington.edu/archis/robust-fractionation.pdf}, 2015.

\bibitem{badri}
H.~Badri, K.~Pitter, E.~Holland, F.~Michor, and K.~Leder.
\newblock Optimization of proneural glioblastoma radiationdosing schedules
  problem.
\newblock {\em Submitted}, 2014.

\bibitem{bentley1979algorithms}
J.~Bentley and T.~Ottmann.
\newblock Algorithms for reporting and counting geometric intersections.
\newblock {\em Computers, IEEE Transactions on}, 100(9):643--647, 1979.

\bibitem{BeMuMc95}
E.~Bernhard, R.~Muschel, and W.~McKenna.
\newblock Effects of ionizing radiation on cell cycle progression.
\newblock {\em Radiation and Environment Biophysics}, 34(2):79--83, 1995.

\bibitem{brenner2008linear}
D.~Brenner.
\newblock The linear-quadratic model is an appropriate methodology for
  determining isoeffective doses at large doses per fraction.
\newblock In {\em Seminars in radiation oncology}, volume~18, pages 234--239.
  Elsevier, 2008.

\bibitem{brenner1998linear}
D.~Brenner, L.~Hlatky, P.~Hahnfeldt, Y.~Huang, and R.~Sachs.
\newblock The linear-quadratic model and most other common radiobiological
  models result in similar predictions of time-dose relationships.
\newblock {\em Radiation research}, 150(1):83--91, 1998.

\bibitem{castadot2010adaptive}
P.~Castadot, J.~Lee, X.~Geets, and V.~Gr{\'e}goire.
\newblock Adaptive radiotherapy of head and neck cancer.
\newblock In {\em Seminars in radiation oncology}, volume~20, pages 84--93.
  Elsevier, 2010.

\bibitem{chan2006robust}
T.~Chan, T.~Bortfeld, and J.~Tsitsiklis.
\newblock A robust approach to imrt optimization.
\newblock {\em Physics in medicine and biology}, 51(10):2567, 2006.

\bibitem{chu2005robust}
M.~Chu, Y.~Zinchenko, S.~Henderson, and M.~Sharpe.
\newblock Robust optimization for intensity modulated radiation therapy
  treatment planning under uncertainty.
\newblock {\em Physics in Medicine and Biology}, 50(23):5463, 2005.

\bibitem{Maciejewski1990oralcavityl}
J.~Denham, Q.~Walker, D.~Lamb, C.~Hamilton, P.~O'Brien, N.~Spry, A.~Hindley,
  M.~Poulsen, M.~O'Brien, and L.~Tripcony.
\newblock Dose fractionation and regeneration in radiotherapy for cancer of the
  oral cavity and oropharynx. part 2. normal tissue responses: acute and late
  effects.
\newblock {\em International Journal of Radiation Oncology* Biology* Physics},
  18(1):101--111, 1990.

\bibitem{emami1991tolerance}
B.~Emami, J.~Lyman, A.~Brown, L.~Cola, M.~Goitein, J.~Munzenrider, B.~Shank,
  L.~Solin, and M.~Wesson.
\newblock Tolerance of normal tissue to therapeutic irradiation.
\newblock {\em International Journal of Radiation Oncology* Biology* Physics},
  21(1):109--122, 1991.

\bibitem{FaWeLi08}
C.~Fakhry, W.~Westra, S.~Li, A.~Cmelak, J.~Ridge, H.~Pinto, A.~Forastiere, and
  M.~Gillison.
\newblock Improved survival of patients with human papillomavirus�positive
  head and neck squamous cell carcinoma in a prospective clinical trial.
\newblock {\em Journal of the National Cancer Institute}, 100:261--269, 2008.

\bibitem{fiorino2006significant}
C.~Fiorino, I.~Dell'Oca, A.~Pierelli, S.~Broggi, E.~Martin, N.~Muzio,
  B.~Longobardi, F.~Fazio, and R.~Calandrino.
\newblock Significant improvement in normal tissue sparing and target coverage
  for head and neck cancer by means of helical tomotherapy.
\newblock {\em Radiotherapy and oncology}, 78(3):276--282, 2006.

\bibitem{FoViAl99}
H.~Forrester, C.~Vidair, N.~Albright, C.~Ling, and W.~Dewey.
\newblock Using computerized video time lapse for quantifying cell death of
  {X}-irradiated rat embryo cells transfected with c-myc or c-ha-ras.
\newblock {\em Cancer Research}, 59:931--939, 1999.

\bibitem{fowler1989linear}
F.~Fowler.
\newblock The linear-quadratic formula and progress in fractionated
  radiotherapy.
\newblock {\em British Journal of Radiology}, 62(740):679--694, 1989.

\bibitem{fowler201421}
J.~Fowler.
\newblock 21 years of biologically effective dose.
\newblock 2014.

\bibitem{hall2006radiobiology}
E.~Hall and A.~Giaccia.
\newblock {\em Radiobiology for the Radiologist}.
\newblock Wolters Kluwer Health, 2006.

\bibitem{HaCo10}
M.~Hamburg and F.~Collins.
\newblock The path to personalized medicine.
\newblock {\em The New England Journal of Medicine}, 363:301--304, 2010.

\bibitem{joiner2009basic}
M.~Joiner and A.~van~der Kogel.
\newblock {\em Basic Clinical Radiobiology Fourth Edition}.
\newblock CRC Press, 2009.

\bibitem{LaErOr12}
K.~Lauber, A.~Ernst, M.~Orth, M.~Hermann, and C.~Belka.
\newblock Dying cell clearance and its impact on the outcome of tumor
  radiotherapy.
\newblock {\em Frontiers in Oncology}, 2, 2012.

\bibitem{leder2014mathematical}
K.~Leder, K.~Pitter, Q.~LaPlant, D.~Hambardzumyan, B.~Ross, T.~Chan,
  E.~Holland, and F.~Michor.
\newblock Mathematical modeling of pdgf-driven glioblastoma reveals optimized
  radiation dosing schedules.
\newblock {\em Cell}, 156(3):603--616, 2014.

\bibitem{lof1995optimal}
J.~Lof, B.~Lind, and A.~Brahme.
\newblock Optimal radiation beam profiles considering the stochastic process of
  patient positioning in fractionated radiation therapy.
\newblock {\em Inverse Problems}, 11(6):1189, 1995.

\bibitem{lof1998adaptive}
J.~L{\"o}f, B.~Lind, and A.~Brahme.
\newblock An adaptive control algorithm for optimization of intensity modulated
  radiotherapy considering uncertainties in beam profiles, patient set-up and
  internal organ motion.
\newblock {\em Physics in medicine and biology}, 43(6):1605, 1998.

\bibitem{maciejewski1986alpha}
B.~Maciejewski, J.~Taylor, and H.~Withers.
\newblock Alpha/beta value and the importance of size of dose per fraction for
  late complications in the supraglottic larynx.
\newblock {\em Radiotherapy and Oncology}, 7(4):323--326, 1986.

\bibitem{meeks2000calculation}
S.~Meeks, J.~Buatti, K.~Foote, W.~Friedman, and F.~Bova.
\newblock Calculation of cranial nerve complication probability for acoustic
  neuroma radiosurgery.
\newblock {\em International Journal of Radiation Oncology* Biology* Physics},
  47(3):597--602, 2000.

\bibitem{MeHaCa14}
A.~Menzies, L.~Haydu, M.~Carlino, M.~Azer, P.~Carr, R.~Kefford, and G.~Long.
\newblock Inter- and intra-patient heterogeneity of response and progression to
  targeted therapy in metastatic melanoma.
\newblock {\em PLoS ONE}, e0085004, 2014.

\bibitem{mizuta2012mathematical}
M.~Mizuta, S.~Takao, H.~Date, N.~Kishimotoi, K.~Sutherland, R.~Onimaru, and
  H.~Shirato.
\newblock A mathematical study to select fractionation regimen based on
  physical dose distribution and the linear--quadratic model.
\newblock {\em International Journal of Radiation Oncology* Biology* Physics},
  84(3):829--833, 2012.

\bibitem{TCGA_GBM}
TCGA~Research Network.
\newblock The somatic genomic landscape of glioblastoma.
\newblock {\em Cell}, 155:462--477, 2013.

\bibitem{TCGA_BC}
The Cancer Genome~Atlas Network.
\newblock Comprehensive molecular portraits of human breast tumours.
\newblock {\em Nature}, 490:61--70, 2012.

\bibitem{orton1988unified}
C.~Orton and L.~Cohen.
\newblock A unified approach to dose-effect relationships in radiotherapy. i:
  Modified tdf and linear quadratic equations.
\newblock {\em International Journal of Radiation Oncology* Biology* Physics},
  14(3):549--556, 1988.

\bibitem{pan2007physical}
C.~Pan, A.~Eisbruch, R.~Ten Haken, et~al.
\newblock Physical models and simpler dosimetric descriptors of radiation late
  toxicity.
\newblock In {\em Seminars in radiation oncology}, volume~17, pages 108--120.
  Elsevier, 2007.

\bibitem{pflugfelder2008worst}
D.~Pflugfelder, J.~Wilkens, and U.~Oelfke.
\newblock Worst case optimization: a method to account for uncertainties in the
  optimization of intensity modulated proton therapy.
\newblock {\em Physics in medicine and biology}, 53(6):1689, 2008.

\bibitem{pierre1969optimization}
D.~Pierre.
\newblock {\em Optimization theory with applications}.
\newblock DoverPublications. com, 1969.

\bibitem{rezvani1993sensitivity}
M.~Rezvani, J.~Fowler, J.~Hopewell, and C.~Alcock.
\newblock Sensitivity of human squamous cell carcinoma of the larynx to
  fractionated radiotherapy.
\newblock {\em The British journal of radiology}, 66(783):245--255, 1993.

\bibitem{ruifrok1992fractionation}
A.~Ruifrok, B.~Kleiboer, and A.~Van der Kogel.
\newblock Fractionation sensitivity of the rat cervical spinal cord during
  radiation retreatment.
\newblock {\em Radiotherapy and Oncology}, 25(4):295--300, 1992.

\bibitem{saberian2014optimal}
F.~Saberian, A.~Ghate, and M.~Kim.
\newblock Optimal fractionation in radiotherapy with multiple normal tissues.
\newblock {\em Available at SSRN 2478481}, 2014.

\bibitem{SaGhKi15}
F.~Saberian, A.~Ghate, and M.~Kim.
\newblock A two-variable linear program solves the standard linear-quadratic
  formulation of the fractionation problem in cancer radiotherapy.
\newblock {\em Operations Research Letters}, 2015.

\bibitem{saibishkumar2007sparing}
E.~Saibishkumar, N.~Jha, R.~Scrimger, M.~MacKenzie, H.~Daly, C.~Field,
  G.~Fallone, and M.~Parliament.
\newblock Sparing the parotid glands and surgically transferred submandibular
  gland with helical tomotherapy in post-operative radiation of head and neck
  cancer: a planning study.
\newblock {\em Radiotherapy and Oncology}, 85(1):98--104, 2007.

\bibitem{stroom1999inclusion}
J.~Stroom, H.~de~Boer, H.~Huizenga, and A.~Visser.
\newblock Inclusion of geometrical uncertainties in radiotherapy treatment
  planning by means of coverage probability.
\newblock {\em International Journal of Radiation Oncology* Biology* Physics},
  43(4):905--919, 1999.

\bibitem{travis1987isoeffect}
E.~Travis and S.~Tucker.
\newblock Isoeffect models and fractionated radiation therapy.
\newblock {\em International Journal of Radiation Oncology* Biology* Physics},
  13(2):283--287, 1987.

\bibitem{turesson1989repair}
I.~Turesson and H.~Thames.
\newblock Repair capacity and kinetics of human skin during fractionated
  radiotherapy: erythema, desquamation, and telangiectasia after 3 and 5 year's
  follow-up.
\newblock {\em Radiotherapy and Oncology}, 15(2):169--188, 1989.

\bibitem{unkelbach2007accounting}
J.~Unkelbach, T.~Chan, and T.~Bortfeld.
\newblock Accounting for range uncertainties in the optimization of intensity
  modulated proton therapy.
\newblock {\em Physics in medicine and biology}, 52(10):2755, 2007.

\bibitem{unkelbach2013dependence}
J.~Unkelbach, D.~Craft, E.~Salari, J.~Ramakrishnan, and T.~Bortfeld.
\newblock The dependence of optimal fractionation schemes on the spatial dose
  distribution.
\newblock {\em Physics in medicine and biology}, 58(1):159, 2013.

\bibitem{unkelbach2004inclusion}
J.~Unkelbach and U.~Oelfke.
\newblock Inclusion of organ movements in imrt treatment planning via inverse
  planning based on probability distributions.
\newblock {\em Physics in medicine and biology}, 49(17):4005, 2004.

\bibitem{withers1975four}
H.~Withers.
\newblock Four r's of radiotherapy.
\newblock {\em Adv. Radiat. Biol., v. 5, pp. 241-247}, 5, 1975.

\bibitem{withers1995late}
H.~Withers, L.~Peters, J.~Taylor, J.~Owen, W.~Morrison, T.~Schultheiss,
  T.~Keane, B.~O'Sullivan, J.~van Dyk, N.~Gupta, et~al.
\newblock Late normal tissue sequelae from radiation therapy for carcinoma of
  the tonsil: patterns of fractionation study of radiobiology.
\newblock {\em International Journal of Radiation Oncology* Biology* Physics},
  33(3):563--568, 1995.

\bibitem{yang2005optimization}
Y.~Yang and L.~Xing.
\newblock Optimization of radiotherapy dose-time fractionation with
  consideration of tumor specific biology.
\newblock {\em Medical physics}, 32:3666, 2005.

\end{thebibliography}
 
\end{document}